\documentclass[12pt,a4paper,final]{ioparta}
\usepackage{graphicx}
\usepackage[breaklinks=true,colorlinks=true,linkcolor=blue,urlcolor=blue,citecolor=blue]{hyperref}
\usepackage{amsmath,amsfonts,amssymb}
\usepackage{slashed}
\usepackage{bm}% bold math
\usepackage[utf8]{inputenc}
\usepackage{cite}  %create interval for many citations at once
\numberwithin{equation}{section}

\begin{document}

\title{Fundamentals of the 3-3-1 Model with Heavy Leptons}

\author{F.C. Correia$^{1,2}$}
\address{$^1$Institute for Theoretical Physics,
	Sao Paulo State University - UNESP.
	Sao Paulo, SP 01140070,
	Brazil.}
\address{$^2$Instituto de Física Corpuscular, Universitat de València. E-46980 Paterna, Spain.}
\ead{ccorreia@ift.unesp.br}

\begin{abstract}
The work is a brief review of the theory based on the $SU(3)_c \otimes SU(3)_L \otimes U(1)_X$ gauge group in the presence of Heavy Leptons. The recent studies in \cite{Cao:2016uur} have established a set of four possible variants for the 3-3-1HL, whose content arises according to the so-denoted variable $\beta$. Since it has been argued about the presence of stable charged particles in this sort of models, we divide the different sectors of the Lagrangian between universal and specific vertices, and 
conclude that the omission of $\beta$-dependent terms in the potential may induce to a discrete symmetry for the versions defined by $|\beta|=\sqrt{3}$. In the context of $|\beta|=\frac{1}{\sqrt{3}}$, where the new degrees of freedom have the same standard electric charges, additional Yukawa interactions may create decay channels into the SM sector. Furthermore, we introduce a method of diagonalization by parts in the Scalar sector motivated by a general consequence of the Goldstone theorem. In summary, we develop the most complete set of terms allowed by the symmetry group and resolve their definitive pieces in order to justify the model description present in the literature.

\end{abstract}


\section{Introduction}
\label{sec:intro}

\newcommand{\covder}{D_\mu}
\newcommand{\NM}{SU(3)_L \otimes U(1)_X}
\newcommand{\SM}{SU(2)_L \otimes U(1)_Y}

The new gauge-structure defining the electroweak sector of 3-3-1 models \cite{Pisano:1991ee}, namely $\NM$, can render important phenomenological consequences and has been object of attention along the last years. Among some recent analysis, we can mention those based on the context of collider \cite{Cao:2016uur} and low-energy physics \cite{Singer:1980sw,Buras:2012dp,Buras:2015kwd,Buras:2016dxz,DeConto:2016osh,Kelso:2014qka}, WIMPs \cite{Mizukoshi:2010ky,Profumo:2013sca} and different possible extensions \cite{DeConto:2015eia}.

The model has a total of six versions, two of which filling the lepton content with a conjugate of standard particles, thus defining the minimal 3-3-1 (see \cite{Fonseca:2016xsy} and references therein). The present work is a brief review of the variant including new Heavy-Leptons \cite{Pleitez:1992xh,Pleitez:1994pu,Ozer:1995xi,Diaz:2003dk}, here denoted like 3-3-1HL. We will see that a different particle content may be defined according to a discrete variable $\beta$ limited to a set of four possible values \cite{Cao:2016uur,Diaz:2004fs}. We then explore the total Lagrangian, divided into boson and fermion sectors, first introducing the totality of $\beta$-independent terms. At this point we aim to conclude about the presence of stable charged particles, as stated in \cite{Cao:2016uur}. 
These features  arise whenever only the universal potential is considered, i.e. when the scalar self-interactions are assumed to be defined exclusively from some generic terms present in all versions of the model. We will see that the omission of $\beta$-specific interactions retain the mixing from the potential following the same pattern as that present in the gauge-fixing Lagrangian. Then we develop the complete case and introduce a method that explore our previous knowledge on the diagonalization of the gauge-dependent mass matrix in order to simplify the search for the eigenstates of the total mass matrix.

The review composes the first part of a work in progress that intends to apply a general integration method for these sort of models, resulting in an Effective Theory that might be directly tested through precision observables. Thus, our first task involves the development of a consistent notation followed by a systematic classification of the interactions among new and standard terms. We ultimately must select and treat those pieces that can generate 6-dimension operators at tree and loop-level. Apart from that, along our analysis we consistently consider one exclusive assumption - The first breaking scale must be much larger than the second one, with the notation translated into $u \gg v_\rho, v_\eta$. 

Finally, we freely follow the same steps of the authors in \cite{CBranco} along their presentation of the Standard Model. Like any gauge theory with spontaneous symmetry breaking, the total Lagrangian is composed by the gauge-kinetic interactions of scalars and fermions, self-interactions of bosons and a Yukawa, such that the review can be organized as follows:

\begin{itemize}
	\item Section \ref{Sec:A} : Gauge Structure and Scalars in the 3-3-1HL
	   \subitem - \ref{Sec:A1} : The Particle Content in Different Versions;
	   \subitem - \ref{Sec:A2} : Self-Interactions of Scalars;
	   \subitem -  \ref{Sec:A3} : Vacuum Stability Condition;
	   \subitem - \ref{Sec:A4} : Gauge-Fixing Lagrangian;
	   \subitem - \ref{Sec:A5} : Gauge-Boson Masses;
	   \subitem - \ref{Sec:A6} : Scalar Masses;
	   \subitem - \ref{Sec:A7} : The Potential for particular models;
	   \subitem - \ref{Sec:A8} : Self-Interactions of Gauge Bosons;
	\item Section \ref{Sec:B} : Fermions in the 3-3-1HL
	   \subitem - \ref{Sec:B1} : Gauge Interactions of the Fermions;
	   \subitem - \ref{Sec:B2} :  Yukawa Lagrangian;
	\item Section \ref{Sec:C} Conclusions.
\end{itemize}

\section{Gauge Structure and Scalars in the 3-3-1HL}\label{Sec:A}
The root of electroweak interactions is expressed by the structure of the covariant derivative. In the context of a $SU(3)_L \otimes U(1)_X$ gauge group it can be represented like:
\begin{equation}
D_\mu = \partial _\mu + i g \mathbf{W}_\mu \cdot \mathbf{I} + i g_X X W_\mu ^0 \mathbb{I} 
\end{equation}
where the bold letter express a simple vector $\mathbf{A} \equiv \left(A^1, A^2, \cdots, A^8\right)$ and $I^a = \frac{\lambda^a}{2}$ are the generators of $SU(3)$. By expanding the gauge piece of $ \covder $ explicitly we can divide it into complex (or non-diagonal) and real interactions
\begin{subequations}
	\begin{equation}\label{Eq:CCint}
	\covder^{(CC)} = i \frac{g}{2} 
	\begin{pmatrix}
	0 & W_\mu^1 - i W_\mu^2 & W_\mu^4 - i W_\mu^5 \\
	W_\mu^1 + i W_\mu^2 & 0 & W_\mu^6 - i W_\mu^7 \\
	W_\mu^4 + i W_\mu^5 & W_\mu^6 + i W_\mu^7 & 0
	\end{pmatrix}
	\end{equation}
	and
	\begin{equation}
	\covder^{(NC)} = i
	\begin{pmatrix}
	\frac{g}{2} \left(W_\mu^3 + \frac{W_\mu^8}{\sqrt{3}}\right) + g_X X W_\mu ^0& 0 & 0 \\
	0 & \frac{g}{2} \left(- W_\mu^3 + \frac{W_\mu^8}{\sqrt{3}}\right) + g_X X W_\mu ^0 & 0 \\
	0 & 0 & - \frac{g}{\sqrt{3}}W_\mu^8 + g_X X W_\mu ^0
	\end{pmatrix}
	\end{equation}
\end{subequations}
In fact, the assertion that the fields presented in Eq.(\ref{Eq:CCint}) will be associated with charged currents would be a first sign on how we will construct both the particle content of the theory and the pattern of symmetry breaking. Nevertheless, we will examine in Section \ref{Sec:A1} that there are in effect two different variants of the model where Neutral Currents may also reside in non-diagonal vertices.

We are also going to consider the following representation to the fermionic fields:
\begin{itemize}
	\item \textbf{Leptons} $\psi_{\alpha}$:  $\begin{pmatrix}
	\left(\nu_\alpha \ l_{\alpha} \right) & E_\alpha
	\end{pmatrix}^\intercal_L$, \qquad $(\mathbf{1},\mathbf{3}, X_\psi)$
	\item \textbf{Quarks} $Q_i$: $\begin{pmatrix}
	\left(d_i \ u_{i}\right) & J_i
	\end{pmatrix}^\intercal_L$, \qquad $(\mathbf{3},\overline{\mathbf{3}}, X_Q)$
	\item \textbf{Quarks} $Q_3$: $\begin{pmatrix}
	\left(u_3 \ d_3 \right) & J_3
	\end{pmatrix}^\intercal_L$, \qquad $(\mathbf{3},\mathbf{3}, X_3)$ 
	\item $u_a^R$: $(\mathbf{3},\mathbf{1}, \frac{2}{3})$
	\item $d_a^R$: $(\mathbf{3},\mathbf{1}, - \frac{1}{3})$
	\item $l_\alpha^R$: $\left(\mathbf{1},\mathbf{1},- 1\right)$ 
	\item $J_i^R$: $(\mathbf{3},\mathbf{1}, X_J)$
	\item $J_3^R$: $(\mathbf{3},\mathbf{1}, X_{J_3})$  
	\item $E^R_\alpha$: $(\mathbf{1},\mathbf{1}, X_E)$
\end{itemize}
where $\alpha = \left[e, \mu, \tau\right]$, $a = \left[1,2,3\right]$ and $i = \left[1,2\right]$. The brackets inside the triplets are denoting that these are in reality doublets of $SU(2)$ or, implicitly, that the gauge group will follow a hierarchy under the symmetry breaking such that, in one of its steps, these objects compose a new symmetric Lagrangian comprising the Standard Model. Apart from that, we omit some of the hypercharges and leave their derivation to the following subsection.

%----------- Chi Triplet ------------------
%\textbf{Scalar} $\chi$:  $\begin{pmatrix}	(\chi^{q_1}, \chi^{q_2}), \tilde{\chi}^0	\end{pmatrix}$, \qquad $(\mathbf{1},\mathbf{3}, X_\chi)$ 

The above mentioned hierarchy is endowing the model with a new scale represented by the letter $u$ and present for the breaking of $\NM$ into $\SM$. The scale is introduced via the vaccuum expectation value of a new neutral scalar $\chi^0$ singlet of $SU(2)$, composing the triplet
$$\chi = \begin{pmatrix}
\left( \chi^V \ \chi^U \right) & \chi^0
\end{pmatrix}^\intercal$$ 

Through the next sections we will discuss that the 3-3-1HL must also include two additional scalar triplets, such that it might also appear in an intermediate scale via Higgs interactions. Nevertheless, for a matter of clarity and driven by a current phenomenology that disallow new degrees of freedom to show up at low energies, we may perform our task of separating the model into standard and new exotic sectors by considering only the first breaking. In other words, we can place the classical electroweak symmetry breaking aside, enclosed in a $\mathcal{L}_{SM}$ function. Thus, we are focused primarily in an universe ruled by nine interactions, where five of them are mediated by massive and four by massless vector particles. The former acquire their masses when $\chi^0$ presents a v.e.v. $\langle\chi^0\rangle = u$, or $\langle\chi\rangle \propto \left(0 \ 0 \ u\right)^\intercal$, leading the theory to $\mathcal{L}_{331} \rightarrow \mathcal{L}_{SM} + \text{NP}$.

Inside the set of nine $\NM$ generators, the following four leave the vacuum unbroken and may define a basis for the group $\SM$ \cite{Buras:2012dp}
\begin{equation}
\mathbb{T}_1 \langle\chi\rangle = \mathbb{T}_2 \langle\chi\rangle = \mathbb{T}_3 \langle\chi\rangle = (\beta \mathbb{T}_8 + X\mathbb{I}) \langle\chi\rangle = 0
\end{equation}
From our knowledge on the SM symmetry breaking, in general we can extract a connection between electromagnetism and weak interactions represented by the so called Gell-Mann-Nishima relation
\begin{subequations}
	\begin{equation}\label{GNR1}
	\mathbb{Q} = \mathbb{T}_3 + \frac{\mathbb{Y}}{2}
	\end{equation} 
	where, in our case,
	\begin{equation}\label{GNR2}
	\frac{\mathbb{Y}}{2} = \beta \mathbb{T}_8 + X\mathbb{I}
	\end{equation}
\end{subequations}
i.e. a diagonal generator for the SM symmetry group defining the particles hypercharge and the respective conserved currents.

The Eq.(\ref{GNR2}) introduces a new and decisive variable. The parameter $\beta$ may completely alter the model phenomenology and is an element of a set limited to four numbers, namely $\bigl\{\pm \sqrt{3}, \pm \frac{1}{\sqrt{3}}\bigr\}$. As presented in \cite{Cao:2016uur}, this set is generated by a series of factors which includes the integer nature of the electric charge for asymptotic states and the positiveness for variables of mass. For illustration, we conclude this section with the demonstration of some results and identities in the context of $\beta = - \sqrt{3}$ (or $X_\chi = -1$). This version contains a single charged heavy lepton and the previous relations may provide us, for example, the charges of exotic quarks inside the triplets.

Since the quarks $J$ are singlet of $SU(2)$, we have
\begin{eqnarray}
Q_3 = \begin{pmatrix}  u_3 \\ d_3 \\ J_3 \end{pmatrix} \qquad \rightarrow \qquad
\mathbb{Q}_{Q_3} = \begin{pmatrix} \frac{2}{3} &  & \\  & -\frac{1}{3} &  \\  &  & q_{J_3} \end{pmatrix}
\end{eqnarray} 
\begin{eqnarray}
\therefore \quad \frac{2}{3} &=& (\mathbb{T}_3)_{11} -\sqrt{3} (\mathbb{T}_8)_{11} + X_3 \nonumber \\
&=& X_3
\end{eqnarray}
The $J_3$ electric charge will then be given by
\begin{equation}
q_{J_3} = -\sqrt{3} (\mathbb{T}_8)_{33} + \frac{2}{3} = \frac{5}{3} 
\end{equation}
We can observe, for instance, that the gauge boson ($W^4 - i W^5$) will couple with the current $$j^\mu_{(uJ)} \equiv \overline{u}_{3L} \gamma^\mu J_{3L}$$ corresponding thus to a particle charged by $(-1)$. Similarly, we have $$j^\mu_{(dJ)} \equiv \overline{d}_{3L} \gamma^\mu J_{3L}$$ coupled to what we are going to denote U-boson and whose charge, in this case, must be equal to $(-2)$.

If we consider the three $SU(2)$ subalgebras of $SU(3)$ defined by the raising and lowering operators:
\begin{eqnarray}
\mathbb{I}_\pm &=& \frac{\mathbb{T}_1 \pm i \mathbb{T}_2}{\sqrt{2}}, \qquad \mathbb{I}_3 = \mathbb{T}_3  \\
\mathbb{J}_\pm &=& \frac{\mathbb{T}_4 \pm i \mathbb{T}_5}{\sqrt{2}}, \qquad \mathbb{J}_3 = \frac{\sqrt{3}}{2} \mathbb{T}_8 - \frac{\mathbb{T}_3}{2} \\
\mathbb{L}_\pm &=& \frac{\mathbb{T}_6 \pm i \mathbb{T}_7}{\sqrt{2}}, \qquad \mathbb{L}_3 = \frac{\sqrt{3}}{2} \mathbb{T}_8 + \frac{\mathbb{T}_3}{2} 
\end{eqnarray}
and also the gauge bosons (note the changes on the signs for $V$ and $U$ compared with $W$):
\begin{eqnarray}
W^\pm = \frac{W^1 \mp i W^2}{\sqrt{2}}, \quad 
V^\pm = \frac{W^4 \pm i W^5}{\sqrt{2}},  \quad
U^{\pm\pm} = \frac{W^6 \pm i W^7}{\sqrt{2}}
\end{eqnarray}
the charged sector of $D_\mu$ can be expressed in a short notation as
\begin{equation}
\covder^{(CC)} = i g (W^+ \mathbb{I}_+ + W^- \mathbb{I}_- + V^- \mathbb{J}_+ + V^+ \mathbb{J}_- + U^{--} \mathbb{L}_+ + U^{++} \mathbb{L}_- )
\end{equation}
In order to obtain the interactions in the conjugate representation we just change the raising and lowering operators sign: 
\begin{equation}
\covder^{*(CC)} = - i g (W^+ \mathbb{I}_- + W^- \mathbb{I}_+ + V^- \mathbb{J}_- + V^+ \mathbb{J}_+ + U^{--} \mathbb{L}_- + U^{++} \mathbb{L}_+ )
\end{equation}
It is common to present the quark representations with a minus sign in the definition of $Q_i$, in general, aiming to recover the exact aspect of the SM Lagrangian. Nevertheless, such phase insertion does not have any phenomenological implications.

\begin{itemize}
	\item \textbf{Note}: To find the Gell-Mann-Nishima relation to the fields in conjugate representation we must remember that the conjugation of the generators $\mathbb{T}_3$ and $\mathbb{T}_8$ is defined as
	\begin{equation}
	\mathbb{T}^*_3 = - \mathbb{T}_3, \qquad \mathbb{T}^*_8 = -\mathbb{T}_8
	\end{equation}
	and, therefore, 
	\begin{equation}\label{GNRC}
	\mathbb{Q} = - \mathbb{T}_3 + \sqrt{3} \mathbb{T}_8 + X\mathbb{I}
	\end{equation}
	Let us find, for example, $X_Q$ and $q_{J_i}$:
	\begin{eqnarray}
	\mathbb{Q}_{Q_i} = \begin{pmatrix} - \frac{1}{3} &  & \\  & \frac{2}{3} &  \\  &  & q_{J_i} \end{pmatrix}
	\quad \rightarrow \quad \frac{2}{3} &=& (- \mathbb{T}_3)_{22} + \sqrt{3} (\mathbb{T}_8)_{22} + X_Q \nonumber \\
	&=& \frac{1}{2} + \frac{1}{2} +  X_Q \nonumber 
	\end{eqnarray}
	Thus, 
	\begin{equation}
	\therefore \qquad X_Q = - \frac{1}{3} \qquad \rightarrow \qquad q_{J_i} = \sqrt{3} (\mathbb{T}_8)_{33} - \frac{1}{3} = -\frac{4}{3} 
	\end{equation}
\end{itemize}

Now we move to the diagonal terms of the covariant derivative, those corresponding to the neutral current interactions:
\begin{equation}
\covder^{(NC)} = i g (\mathbb{T}_3 W^3 +  \mathbb{T}_8 W^8) + i g_X X \mathbb{I} W^0
\end{equation}
After the first breaking by $\langle \chi \rangle$, $W^0$ will mix with $W^8$ and we can rotate them into the mass eigenstates by
\begin{equation}
\begin{pmatrix}
W^0 \\ W^8
\end{pmatrix} = \begin{pmatrix}
c_x & -s_x \\ s_x & c_x
\end{pmatrix} 
\begin{pmatrix}
B \\
Z'
\end{pmatrix}
\end{equation}  
which implies
\begin{eqnarray}
\covder^{(NC)} &=& i g \mathbb{T}_3 W^3 +  ig \mathbb{T}_8 (B s_x + Z' c_x) + i g_X X \mathbb{I} (B c_x - Z' s_x) \nonumber \\
&=& i (\mathbf{g}_{W^3} W^3 +  \mathbf{g}_B B + \mathbf{g}_{Z'} Z')
\end{eqnarray}
where we have defined
\begin{equation}
\mathbf{g}_{W^3} \equiv g \mathbb{T}_3, \qquad  
\mathbf{g}_B \equiv g \mathbb{T}_8 s_x + g_X X \mathbb{I} c_x, \qquad
\mathbf{g}_{Z'} \equiv \mathbb{T}_8 c_x - g_X X \mathbb{I} s_x
\end{equation}
We can also anticipate the mixing between $B$ and $W^3$ after the second symmetry breaking
\begin{equation}
\begin{pmatrix}
B \\ W^3
\end{pmatrix} = \begin{pmatrix}
c_w & -s_w \\ s_w & c_w
\end{pmatrix} 
\begin{pmatrix}
A \\
Z
\end{pmatrix}
\end{equation}  
or, finally, 
\begin{eqnarray}
\covder^{(NC)} = i (\mathbf{g}_{A} A +  \mathbf{g}_Z Z + \mathbf{g}_{Z'} Z')
\end{eqnarray}
where,
\begin{equation}
\mathbf{g}_{A} \equiv s_w \mathbf{g}_{W^3}  + c_w \mathbf{g}_B , 
\qquad
\mathbf{g}_Z \equiv c_w \mathbf{g}_{W^3} - s_w \mathbf{g}_B 
\end{equation}
In order to require that electromagnetic interactions be reproduced, we must impose
\begin{equation}
\mathbf{g}_{A} = e \mathbb{Q}
\end{equation}
or 
\begin{equation}\label{charge}
e \mathbb{Q} = g \mathbb{T}_3 s_w + g \mathbb{T}_8 s_x c_w + g_X X \mathbb{I} c_x c_w
\end{equation}

The Eq.(\ref{charge}) when applied for different quarks and leptons may result in some important relations to the coupling constants. For instance, to the quark-$d_1$ ($q_d = -\frac{1}{3}, X_Q = -\frac{1}{3}$) we have:
\begin{equation}
-\frac{e}{3} = - \left(\frac{g}{2}s_w + \frac{g}{2\sqrt{3}}s_x c_w\right) - \frac{g_X}{3} c_x c_w  
\end{equation}
From the following Eq.(\ref{sen}) and by assuming $e = g s_w$:
$$-\frac{e}{3} = - \left(\frac{e}{2} - \frac{e}{2} \right) - \frac{g_X}{3} c_x c_w \quad \rightarrow \quad g_X c_x c_w = e  $$ 
or, finally, if $g' = \frac{e}{c_w}$,
\begin{equation}\label{hyp}
g_X c_x = g'
\end{equation}
which connects the hypercharge sectors of the Standard Model and the 3-3-1HL. Besides, from the results of the next section, derived for a generic version of the model, it can be shown that the Eq.(\ref{hyp}) is in fact independent of $\beta$. We must note, however, that we have considered the same relations raised in the context of the Standard Model, that is to assert both angles and couplings as equivalent.  

If we repeat our use of Eq. (\ref{charge}) to the case of neutrinos ($q_\nu = 0, X_\psi = 0$) we find
\begin{equation}\label{sen}
\frac{g}{2} s_w + \frac{g}{2\sqrt{3}} s_x c_w = 0 \qquad \rightarrow \qquad s_x = -\sqrt{3} \tan \theta_w \qquad (\text{or} \ s_x = \beta \tan \theta_w )
\end{equation}
which provides a connection between the Weinberg angle and the mixing on the first breaking. From the identities proved in the next section it is straightforward to obtain the general formula inside the brackets.

We emphasize that the rotation of ($B, W^3$) into ($A, Z$) is in reality neglecting the mixing with $Z'$ emerged after the second breaking. This assumption is also motivated by the large difference between the scales, $u \gg v$.

The Eqs. (\ref{sen}) and (\ref{hyp}) will also induce a set of simplified expressions for the $Z$ and $Z'$ couplings. We write, for instance, 
\begin{eqnarray} \label{Zcoup}
\mathbf{g}_Z &=&  g \mathbb{T}_3 c_w - (g \mathbb{T}_8 s_x + g_X X \mathbb{I} c_x)s_w \nonumber \\
&\stackrel{(\ref{hyp})}{=}& g \mathbb{T}_3 c_w - (g \mathbb{T}_8 s_x + g' X \mathbb{I})s_w \nonumber \\
&\stackrel{(\ref{sen})}{=}& g c_w \mathbb{T}_3 + \sqrt{3} g  \tan \theta_w s_w \mathbb{T}_8 - g's_w X \mathbb{I} \nonumber \\
&=& g c_w \mathbb{T}_3 + \sqrt{3} g  \frac{s^2_w}{c_w} \mathbb{T}_8 - g \frac{s^2_w}{c_w} X \mathbb{I} \nonumber \\
&=& g c_w \mathbb{T}_3 + g  \frac{s^2_w}{c_w}(\sqrt{3} \mathbb{T}_8 - X \mathbb{I}) \nonumber \\
&=& g c_w \mathbb{T}_3 + g  \frac{s^2_w}{c_w}(\mathbb{T}_3 - \mathbb{Q}) \nonumber \\
&=& \frac{g}{c_w} (\mathbb{T}_3 - \mathbb{Q} s^2_w)
\end{eqnarray}
i.e. the well-known coupling of $Z$ with the SM particles. The fourth and sixth equality sign consider $g' = g \tan \theta_w$ and Eq.(\ref{GNRC}), respectively, and the identity is $\beta$-independent. A similar expression can also be achieved for $Z'$ through
\begin{eqnarray} \label{Zpcoup}
\mathbf{g}_{Z'} &=&  g \mathbb{T}_8 c_x - g_X X \mathbb{I} s_x \nonumber \\
&\stackrel{(\ref{hyp})}{=}& g \mathbb{T}_8 c_x - \frac{g'}{c_x} X \mathbb{I} s_x \nonumber \\
&=& g (\mathbb{T}_8 c_x - \tan \theta_w \tan \theta_x X \mathbb{I})\nonumber \\
&\stackrel{(\ref{sen})}{=}& \frac{g}{c_x} (\mathbb{T}_8 c^2_x + \sqrt{3} \tan^2 \theta_w X \mathbb{I}) 
\end{eqnarray}
We note that the relation (\ref{Zcoup}) exhibit the massless nature of $Z$ after the first breaking $$\frac{g}{c_w} ((\mathbb{T}_3)_{33} - 0 * s^2_w) = 0$$ i.e. there is no coupling with the neutral scalar in $\chi$.

Thus, the section may be concluded with the total covariant derivative expressed under a simplified notation,
\begin{eqnarray}\label{covdev}
\covder &=&  \mathbb{I} \partial_\mu + i [g(W^+ \mathbb{I}_+ + W^- \mathbb{I}_-) + \mathbf{g}_Z Z + e \mathbb{Q} A] + \nonumber \\
& & + i [g (V^- \mathbb{J}_+ + V^+ \mathbb{J}_- + U^{--} \mathbb{L}_+ + U^{++} \mathbb{L}_-) + \mathbf{g}_{Z'} Z']
\end{eqnarray}
The first line reproduces exactly the Standard Model contribution and will make simpler the task of dividing the 3-3-1HL into SM and New Physics elements.

As mentioned previously, in this work we discuss how the different sectors of the Lagrangian will connect the new degrees of freedom with the standard fields at tree- and loop-level. In order to be consistent and provide an independent review, we remake in detail some of the main aspects of the model, being however substantially supported by previous works like \cite{Cao:2016uur}, \cite{Pisano:1991ee} and  \cite{Buras:2012dp}. When the context allows, we persist regarding the first breaking $\NM \rightarrow \SM$ with priority. In the next section we extract some identities for a general 3-3-1HL version.

\subsection{Particle Content in Different Versions}\label{Sec:A1}

First, let us recall the electric charge operator:
\begin{equation}\label{GNR}
\mathbb{Q} = \mathbb{T}_3 +  \beta \mathbb{T}_8 + X\mathbb{I}
\end{equation}
Then we start to find the electric charges dependent on $\beta$. From the first line of $D_\mu \chi$, for example, we must have $q_{\chi_1} = 1 + q_{\chi_2} = q_V$, which can also justify why we will define the first entry of $\chi$ as $\chi_V$. Besides, the second line would give us $q_U = q_{\chi_2}$, and we denote $\chi_2 \equiv \chi_U$. 

The neutral third component of $\chi$ implies $X_\chi = \frac{\beta}{\sqrt{3}}$, i.e.

\begin{eqnarray} \label{Eq:chargeV}
q_{\chi_V} = q_V &=& \frac{1}{2} + \frac{\beta}{2 \sqrt{3}} + \frac{\beta}{\sqrt{3}} \nonumber \\
&=& \frac{\sqrt{3}}{2}\beta + \frac{1}{2} \qquad \rightarrow \qquad q_U = \frac{\sqrt{3}}{2}\beta - \frac{1}{2}
\end{eqnarray}

By repeating this simple procedure to the quarks and leptons representations we can obtain their respective hypercharges. For example, the value of $X_Q$, defined in conjugate representation, will be given by
\begin{eqnarray}\label{Eq:qcharge}
X_Q = \frac{1}{6} + \beta \frac{1}{2 \sqrt{3}} \quad \rightarrow \quad q_{j_i} = \frac{1}{6} + \frac{\sqrt{3}}{2}\beta
\end{eqnarray}
or, for the triplet in fundamental representation,
\begin{eqnarray}\label{Eq:qcharge3}
X_3 = \frac{1}{6} - \frac{\beta}{2 \sqrt{3}} \quad \rightarrow \quad q_{j_3} = \frac{1}{6} - \frac{\sqrt{3}}{2}\beta
\end{eqnarray}

As mentioned before, the electric charges of the new gauge bosons, as well as the $Z'$ mass, prevent $\beta$ of assuming any possible value  \cite{Cao:2016uur}, restricting it to the set
\begin{equation}
\beta \in \left[\pm \frac{1}{\sqrt{3}}, \pm \sqrt{3}\right]
\end{equation}
We see from Eq.(\ref{Eq:qcharge}) and Eq.(\ref{Eq:qcharge3}) that the effect of a $\pm$ sign for quarks is merely of inverting the electric charges into different representations. For $|\beta| = \sqrt{3}$ the possible values are $q_J \in \left[-\frac{4}{3}, \frac{5}{3}\right]$ and for $|\beta| = \frac{1}{\sqrt{3}}$, $q_J \in \left[-\frac{1}{3}, \frac{2}{3}\right]$.

Similarly, for the leptons we have:
\begin{equation}
\psi_{\alpha}: \begin{pmatrix}
\nu_\alpha \\ l_{\alpha} \\ E_\alpha
\end{pmatrix}_L \qquad \rightarrow \qquad X_\psi = - \left(\frac{1}{2} + \frac{\beta}{2 \sqrt{3}} \right)
\end{equation}
or
\begin{equation}\label{Eq:lcharge}
q_E = - \biggl(\frac{1}{2} + \frac{\sqrt{3}}{2}\beta \biggr)
\end{equation}
Thus, depending on the selected $\beta$ among the four possible values, the model may include from neutral to doubled charged heavy leptons: 
\begin{equation}\label{Eq:lepcharge}
q_E = \left[-2, -1, 0, +1\right] \quad \text{for} \quad \beta = \left[\sqrt{3}, \frac{1}{\sqrt{3}}, -\frac{1}{\sqrt{3}}, - \sqrt{3}\right]
\end{equation}

From phenomenological reasons we are also interested in heavy neutral particles. The Section \ref{Sec:A2} resolve, in the context of Scalars, that the most general potential can be formulated regardless the sign of $\beta$, i.e. it is sufficient to consider each of the two possible modulus. Thus, in the remaining of this work we probe those models where $\beta = \bigl[- \sqrt{3}, - \frac{1}{\sqrt{3}}\bigr]$. 

Finally, let us summarize the general hypercharge of triplets and anti-triplets:
\begin{itemize}
	\item \textbf{Leptons} $\psi_{\alpha}$: \quad $\left(\mathbf{1},\mathbf{3}, - \left(\frac{1}{2} + \frac{\beta}{2 \sqrt{3}}\right)\right)$
	\item \textbf{Quarks} $Q_i$: \quad $\left(\mathbf{3},\overline{\mathbf{3}},\frac{1}{6} + \frac{\beta}{2 \sqrt{3}}\right)$
	\item \textbf{Quarks} $Q_3$: \quad $\left( \mathbf{3},\mathbf{3},\frac{1}{6} - \frac{\beta}{2 \sqrt{3}}\right)$ 
	\item $u_a^R$: $\left(\mathbf{3},\mathbf{1}, \frac{2}{3}  \right)$
	\item $d_a^R$: $\left(\mathbf{3},\mathbf{1}, - \frac{1}{3} \right)$
	\item $J_i^R$: $\left(\mathbf{3},\mathbf{1}, \frac{1}{6} + \beta \frac{\sqrt{3}}{2} \right)$
	\item $J_3^R$: $\left(\mathbf{3},\mathbf{1}, \frac{1}{6} - \beta \frac{\sqrt{3}}{2}\right)$  
	\item $l_\alpha^R$: $\left(\mathbf{1},\mathbf{1},- 1\right)$
	\item $E_\alpha^R$: $\left(\mathbf{1},\mathbf{1},- \left(\frac{1}{2} + \beta \frac{\sqrt{3}}{2}\right) \right)$
\end{itemize}
where $\alpha = \left[e, \mu, \tau\right]$, $a = \left[1,2,3\right]$ and $i = \left[1,2\right]$. 

We conclude this section with a brief comment on the Scalar triplets. The gauge-fixing Lagrangian, discussed in Section \ref{Sec:A4}, in general is defined from the kinetic sector of the Scalars and contain terms proportional to their product with a correspondent gauge boson. For example, the first line of the product $D^\mu \chi$ contains the sum $\partial^\mu \chi_V + u V$ such that its squared will produce the bilinear $u V \partial \chi_V$. Therefore, we introduce the following notation intending to leave explicit how the gauge-structure of the model will connect the Scalars with Vector Bosons: 
\begin{subequations}\label{Eq:Scadef}
	\begin{eqnarray}
	\chi &=& \begin{pmatrix}
	\chi^{V} \\ \chi^{U} \\ \frac{u + \overline{H} + i \chi_g}{\sqrt{2}}
	\end{pmatrix}, \qquad \left( \mathbf{1},\mathbf{3}, \frac{\beta}{\sqrt{3}} \right) \\ 
	\rho  &=&	\begin{pmatrix}
	\rho^{W} \\ \frac{v_\rho + H_\rho + i \rho_g}{\sqrt{2}} \\  \rho^{-U}
	\end{pmatrix}, \qquad \left( \mathbf{1},\mathbf{3}, \frac{1}{2} - \frac{\beta}{2 \sqrt{3}} \right) \\
	\eta  	&=&
	\begin{pmatrix}
	\frac{v_\eta + H_\eta + i \eta_g}{\sqrt{2}} \\ \eta^{-W} \\  \eta^{-V}
	\end{pmatrix}, \qquad \left( \mathbf{1},\mathbf{3},- \left( \frac{1}{2} + \frac{\beta}{2 \sqrt{3}}\right) \right)
	\end{eqnarray}
\end{subequations}
where the minus sign indicates that their charges are in opposite sign as those defined in Eq.(\ref{Eq:chargeV}).
As we will repeatedly explore, when these connections are not broken by $\beta$-dependent terms, they may control the pattern of scalar mixing and correspond to residual symmetries that avoid the heavy leptons to decay. These features, along with their phenomenological consequences, have been addressed by the authors in \cite{Cao:2016uur}, and here we intend to complement it by exposing the vertices allowed by the gauge symmetry that could create decaying channels into the light asymptotic fields. Thus, we will consider the complete set of elements for the potential and not assume a $Z_2$ symmetry a priori. We demonstrate that the pattern of scalar mixing present in the incomplete and universal\footnote{By `universal' we refer to the components present in all versions of the model.} potential eliminates the totality of tree-level interactions, leaving the heavy-lepton stable.

\subsection{Self-Interactions of Scalars}\label{Sec:A2}
Here we remark and examine the algebraic path followed by the symmetry breaking of a generic potential which may conduce the gauge theories to their correct spectra. We can start then by presenting some aspects of this sector in the Standard Model, where:
\begin{equation}\label{Eq:SMpot}
V(\Phi) = \mu \Phi^\dagger \Phi + \lambda (\Phi^\dagger \Phi)^2
\end{equation}
and
\begin{equation}\label{higgs}
\Phi = \begin{pmatrix}
\varphi^+_g &
\frac{\varphi^0 + i \varphi_g^0}{\sqrt{2}}
\end{pmatrix}^\intercal
\end{equation}

The subindex `g' on some of the above fields is referring to the Goldstone bosons. In fact, an intermediate step along the development of the Goldstone theorem is supported by the condition that the vacuum of the theory is located on the point where one of the neutral fields, namely $\varphi^0$, assume a value different from zero. This Vacuum Stability Condition (VSC) will correspond to one or a set of equations that, when applied back to Eq.(\ref{Eq:SMpot}), cancel the mass terms for these `g' degrees of freedom. The contribution to the mass of this particles will be gauge-dependent and comes from the insertion of a gauge-fixing Lagrangian. That is what we called the path for identifying these massless particles, effectively traced by the theorem.

The vacuum expectation value (v.e.v) is usually denoted by $v$, such that the vacuum condition can be read like
\begin{equation}
\frac{\partial V(\Phi)}{\partial \varphi^0} \Big|_{\langle \varphi^0 \rangle = v } = 0
\end{equation}  
The minimum point represented as $\langle \varphi^0 \rangle = v$ denotes that the derivative is being taken on the fields vacuum expectation value.

We can expand $\varphi^0$ around its v.e.v, $\varphi^0 = v + H$, and rewrite the above condition:
\begin{equation}\label{vac}
\frac{\partial H}{\partial \phi^0} \frac{\partial V}{\partial H}\Big|_{\langle H \rangle = 0 } = 0, \qquad \text{or, shortly,} \qquad \partial_H V\Big|_{H=0} = 0  
\end{equation} 
The potential $V(H)$ is a polynomial and the Eq.(\ref{vac}) implies, therefore, that its linear term on $H$ must vanish, or
\begin{equation}
\mu + \lambda v^2 = 0
\end{equation}
which enables us to rewrite Eq.(\ref{Eq:SMpot}) like
\begin{equation}
V(\Phi) = \lambda \left(\Phi^\dagger \Phi - \frac{v^2}{2}\right)^2
\end{equation}
Thus, the zeroth order term vanishes inside the brackets, leaving only the Higgs with a mass term in the function.

At this point we insist on some important remarks. The VSC defines a set of equations which will give us the correct mass matrix coming from the potential. This is what we see in the simplest one dimensional case of the SM. The rotation to the mass eigenstates must also include the contribution coming from quadratic mixed terms present in the gauge-kinetic scalar Lagrangian. These terms are strictly correlated to the gauge-fixing piece, whose treatment leads to the correct mass of the Goldstone bosons. Finally, the Goldstone theorem is essentially based on the vacuum stability and reveal, as a corollary, that these mass matrices, i.e. those coming from the potential and the gauge-fixing, must reside in orthogonal subspaces. On what follows, we explore this result in detail.

Once we are focusing on the potential, it might be important to consider the whole spectra of scalar fields. In order to the standard quarks and leptons acquire their masses, we introduce two additional triplets with the following components:
\begin{eqnarray}
\beta = \ - \sqrt{3}: \ \rho = \begin{pmatrix} \Phi_\rho \\ \rho^{++}_U \end{pmatrix} \quad (\mathbf{1},\mathbf{3}, 1), \quad 
\eta = \begin{pmatrix} \tilde{\Phi}_\eta \\ \eta_V^+ \end{pmatrix} \quad (\mathbf{1},\mathbf{3}, 0) 
\end{eqnarray} 
where $\Phi_\rho$ and $\tilde{\Phi}_\eta$ are just denoting that these fields have a similar structure to the doublet in Eq.(\ref{higgs}) and its conjugate. The expansion of $\chi^0$ around its v.e.v, $\chi^0 = u + \overline{H}$, introduces the new heavy Higgs, $\overline{H}$. As a matter of counting, we are going to identify those eight scalar fields which are connected, as additional degrees of freedom, to the vectors $V^{\pm}, U^{\pm\pm}, W^\pm, Z'$ and $Z$.

In Section (\ref{Sec:A4}) we emphasize that the choice of a specific gauge can be made independently for each gauge interaction. Through their couplings with the SM gauge bosons, the fields in $\Phi_\rho$ and $\tilde{\Phi}_\eta$ settle a pattern of mixing that will be followed by the composition of the potential presented below. Thus, it may be more elucidative to keep developing the model under an arbitrary gauge. We define, for $\beta = - \sqrt{3}$,
\begin{eqnarray}
\rho = \begin{pmatrix} \rho_W^+ \\ \frac{(v_\rho + H_\rho) + i \rho_g}{\sqrt{2}} \\ \rho_U^{++} \end{pmatrix}, \qquad  \qquad
\eta = \begin{pmatrix} \frac{(v_\eta + H_\eta) + i\eta_g}{\sqrt{2}} \\ \eta_W^- \\ \eta_V^+ \end{pmatrix}
\end{eqnarray} 
and finally introduce the most general $\beta$-independent potential  \cite{Buras:2012dp}
\begin{eqnarray}\label{pot}
V(\chi, \rho, \eta) &=& \mu_\chi \chi^\dagger \chi + \lambda_\chi (\chi^\dagger \chi)^2 + \mu_\eta \eta^\dagger \eta + \lambda_\eta (\eta^\dagger \eta)^2 + \mu_\rho \rho^\dagger \rho + \lambda_\rho (\rho^\dagger \rho)^2 + \nonumber \\
& & + \lambda_{\rho \eta} (\rho^\dagger \rho)(\eta^\dagger \eta) + \lambda_{\rho \chi} (\rho^\dagger \rho)(\chi^\dagger \chi) + \lambda_{\eta \chi} (\eta^\dagger \eta)(\chi^\dagger \chi) + \nonumber \\
& & + \overline{\lambda}_{\rho \eta} (\rho^\dagger \eta)(\eta^\dagger \rho) + \overline{\lambda}_{\rho \chi} (\rho^\dagger \chi)(\chi^\dagger \rho) + \overline{\lambda}_{\eta \chi} (\eta^\dagger \chi)(\chi^\dagger \eta) + \nonumber \\
& & + \sqrt{2} \zeta (\epsilon_{ijk} \rho^i \eta^j \chi^k + \text{h.c.})
\end{eqnarray}
The first two entries of each triplet define a $SU(2)$ doublet, such that any of the above terms are $SU(3), SU(2), U(1)_X, U(1)_Y$ invariant. In addition, from the definitions of Eq.(\ref{Eq:Scadef}), there can also exist a set of new terms added to Eq.(\ref{pot}), in the context of specific values for $\beta$. In other words, depending on the specific model we are regarding, new elements might emerge to the potential.   

\subsection{Vacuum Stability Condition}
\label{Sec:A3}
In this framework, the vacuum stability condition is given by
\begin{equation}
\frac{\partial V(\chi^0, \rho^0, \eta^0)}{\partial \chi^0} \Big|_{\langle \chi^0 \rangle = u} = 0, \quad
\frac{\partial V(\chi^0, \rho^0, \eta^0)}{\partial \rho^0} \Big|_{\langle \rho^0 \rangle = v_\rho} = 0, \quad
\frac{\partial V(\chi^0, \rho^0, \eta^0)}{\partial \eta^0} \Big|_{\langle \eta^0 \rangle = v_\eta} = 0
\end{equation}  
where by $\rho^0$ and $\eta^0$ we denote the neutral component of the respective triplet. The relations above can be translated into
\begin{equation}\label{vstotal}
\partial_{\overline{H}} V(\overline{H}, H_\rho, H_\eta) \Big|_{\overline{H} = 0} = 0, \quad
\partial_{H_\rho} V(\overline{H}, H_\rho, H_\eta) \Big|_{H_\rho = 0} = 0, \quad
\partial_{H_\eta} V(\overline{H}, H_\rho, H_\eta) \Big|_{H_\eta = 0} = 0
\end{equation}  
i.e. the coefficient of linear terms must vanish. For illustration we can draw a geometrical picture up to the potential vacuum. First, let us rewrite Eq.(\ref{pot}) as
\begin{eqnarray}\label{pot2}
V(\chi, \rho, \eta) &=& V|_{\rho, \eta = 0} + V|_{\chi, \eta = 0} + V|_{\chi, \rho = 0} + \nonumber \\
& & + \lambda_{\rho \eta} (\rho^\dagger \rho)(\eta^\dagger \eta) + \lambda_{\rho \chi} (\rho^\dagger \rho)(\chi^\dagger \chi) + \lambda_{\eta \chi} (\eta^\dagger \eta)(\chi^\dagger \chi) + \nonumber \\
& & +  \overline{\lambda}_{\rho \eta} (\rho^\dagger \eta)(\eta^\dagger \rho) + \overline{\lambda}_{\rho \chi} (\rho^\dagger \chi)(\chi^\dagger \rho) + \overline{\lambda}_{\eta \chi} (\eta^\dagger \chi)(\chi^\dagger \eta) + \nonumber \\
& & + \sqrt{2} \zeta (\epsilon_{ijk} \rho^i \eta^j \chi^k + \text{h.c.})
\end{eqnarray}
The first line of Eq.(\ref{pot2}) is composed by the three functions defining the scenario where the symmetry breaking occur independently, i.e. when just one of the triplets are present in the theory. The $V|_{\rho, \eta = 0}$, for instance, comprises the piece regarding solely the breaking $\NM \rightarrow \SM$. The vacuum condition to it provides
\begin{equation}\label{Eq:vscchi}
\partial_{\overline{H}} \left( V|_{\rho, \eta = 0}\right) \Big|_{\overline{H} = 0} = 0 \quad \rightarrow \quad \mu_\chi + \lambda_\chi u^2 = 0
\end{equation}  

Now, when the VSC is also assumed for the second and third terms of the first line, we find
\begin{subequations}\label{Eq:vsrn}
	\begin{eqnarray}
	\partial_{H_\rho} \left( V|_{\chi, \eta = 0}\right) \Big|_{H_\rho = 0} &=& 0 \quad \rightarrow \quad \mu_\rho + \lambda_\rho v_\rho^2 = 0 \\
	\partial_{H_\eta} \left( V|_{\chi, \rho = 0}\right) \Big|_{H_\eta = 0} &=& 0 \quad \rightarrow \quad \mu_\eta + \lambda_\eta v_\eta^2 = 0
	\end{eqnarray}  	
\end{subequations}

When both Eq.(\ref{Eq:vscchi}) and Eq.(\ref{Eq:vsrn}) are taken, we are in fact assuming that the points $(\chi^0, \rho^0, \varphi^0) = \left[(u,0,0), (0,v_\rho,0),(0,0,v_\eta)\right]$ are either local or global minimum of $V(\chi, \rho, \eta)$, along with the point $(u,v_\rho,v_\eta)$. Thus, the vacuum could be described by the following ellipsoid 
\begin{equation}
x^2 + y^2 + z^2 + xy + xz + yz - \frac{(e^2 + u^2)}{u}x - \frac{(e^2 + v_\rho ^2)}{v_\rho} y - \frac{(e^2 + v_\eta ^2)}{v_\eta} z + e^2 = 0
\end{equation}
where $e^2 = \frac{uv_\rho + v_\rho v_\eta + uv_\eta}{2}$ and $(x,y,z) \in \left[(u,0,0), (0,v_\rho,0),(0,0,v_\eta),(u,v_\rho,v_\eta)\right]$ defines a subset of possible solutions.

Finally, by applying Eq.(\ref{vstotal}) simultaneously to the total potential we are left with the ultimate equations
\begin{subequations}\label{Eq:vsto}
	\begin{equation}\label{vstochi}
	\partial_{\overline{H}} V\Big|_{\overline{H} = 0} = 0 \quad \rightarrow \quad \mu_\chi + \lambda_\chi u^2 + \frac{v_\eta ^2}{2}\lambda_{\eta \chi} + \frac{v_\rho ^2}{2} \lambda_{\rho \chi} - \frac{v_\rho v_\eta}{u} \zeta = 0
	\end{equation}  
	\begin{equation}\label{vstrho}
	\partial_{H_\rho} V \Big|_{H_\rho = 0} = 0 \quad \rightarrow \quad \mu_\rho + \lambda_\rho v_\rho ^2 + \frac{u^2}{2} \lambda_{\rho \chi} + \frac{v_\eta ^2}{2} \lambda_{\rho \eta} - \frac{u v_\eta }{v_\rho}\zeta = 0
	\end{equation} 
	and
	\begin{equation}\label{vsteta}
	\partial_{H_\eta} V \Big|_{H_\eta = 0} = 0 \quad \rightarrow \quad \mu_\eta + \lambda_\eta v_\eta ^2 + \frac{u^2}{2} \lambda_{\eta \chi} + \frac{v_\rho ^2}{2} \lambda_{\rho \eta} - \frac{u v_\rho }{v_\eta}\zeta = 0
	\end{equation} 	
\end{subequations}
We are going to see in detail that these three conditions are sufficient to leave the model with a well defined mass spectrum to the scalars.

\subsection{Gauge-Fixing Lagrangian}\label{Sec:A4}
The gauge interactions of the scalars come from the invariant kinetic term
\begin{equation}\label{Dchi}
\mathcal{L}_s \supset (D_\mu \chi)^\dagger(D^\mu \chi)
\end{equation}
For $\beta = - \sqrt{3}$, the scalar can be defined as
\begin{equation}
\chi = \begin{pmatrix}
(\chi_g^- \ \chi_g^{--}) & \frac{\chi_u^0 + i \chi_g}{\sqrt{2}}
\end{pmatrix}^\intercal
\end{equation}
and the $g$ index emphasizes that those states are exactly the Goldstone bosons in the context of a single breaking scale. By applying the Gell-Mann-Nishima relation, Eq.(\ref{GNR1}), we can promptly check that $X_\chi = -1$. 

The covariant derivative (\ref{covdev}), in matrix form, is given by
\begin{eqnarray}\label{CDscalars}
\covder = & \mathbb{I}\partial_\mu 
+ i \frac{g}{\sqrt{2}}\begin{pmatrix}
0 & W^+ & V^-\\
W^- & 0 & U^{--}\\
V^+ & U^{++} & 0 \\
\end{pmatrix} \nonumber \\
& + i \ \text{diag} \begin{bmatrix}
\begin{pmatrix}
-e & \frac{g(\frac{1}{2}+s^2_w)}{c_w} & g \left(\frac{c_x}{2 \sqrt{3}} - \sqrt{3} \text{tg}_w \text{tg}_x\right) \\
-2e & \frac{g(-\frac{1}{2} + 2s^2_w)}{c_w} & g \left(\frac{c_x}{2 \sqrt{3}} - \sqrt{3} \text{tg}_w \text{tg}_x\right) \\
0 & 0 & g \left(- \frac{c_x}{\sqrt{3}} - \sqrt{3} \text{tg}_w \text{tg}_x\right) \\
\end{pmatrix}
\begin{pmatrix}
A \\
Z \\
Z' \\
\end{pmatrix} 
\end{bmatrix}
\end{eqnarray}
where by `diag' we require the vector inside the brackets to define a diagonal $3 \times 3$ matrix. Here, we finish our insistent discussion about the CD - The Eq. (\ref{covdev}) provides a direct representation whence we can see how the gauge interactions play with the isospin components of the triplets, aside from the manifest separation into the SM and New Physics terms, one of the main purposes of this work. Nevertheless, in this section we search for the complete vertices concerning scalars and vectors, and the expansion in Eq. (\ref{CDscalars}) can make their identification simpler.

From the complete result of Eq.(\ref{Dchi}) we can select only the terms that will be indeed related in our analysis. First, we must include those canceled out by the gauge-fixing Lagrangian. Since each factor in $\covder \chi$ is at least linear in the fields, we set up the quadratic parts plus all the $\chi_u^0$ dependent ones, which yield mass to the gauge bosons and introduce the new physical Higgs $\overline{H}$. This corresponds to
\begin{eqnarray}
 (D_\mu \chi)^\dagger(D^\mu \chi) &=& \frac{1}{2}(\partial \chi_u^0 - i \partial \chi^0_g - i n_{33} \chi_u^0 Z')(\partial \chi_u^0 + i \partial \chi^0_g + i n_{33} \chi_u^0 Z') + \nonumber \\
& & + (\partial \chi_g^- + i \frac{g}{2} \chi_u^0 V^-)(\partial \chi_g^+ - i \frac{g}{2} \chi_u^0 V^+) + \nonumber \\
& & + (\partial \chi_g^{--} 
+ i \frac{g}{2} \chi_u^0 U^{--})(\partial \chi_g^{++} - i \frac{g}{2} \chi_u^0 U^{++}) + \nonumber \\
& & + \mathcal{O}(\text{non-quadratic terms}) \qquad (\beta = - \sqrt{3})
\end{eqnarray}
where $n_{33} \equiv g \left(- \frac{c_x}{\sqrt{3}} - \sqrt{3} \text{tg}_w \text{tg}_x\right)$. 
Our final vertices will be given by:
\begin{eqnarray}\label{Eq:selfintchi}
(D_\mu \chi)^\dagger(D^\mu \chi) =& (\partial \chi_u^0)(\partial \chi_u^0) + (\partial \chi^0_g)(\partial \chi^0_g) + \nonumber \\
& + (\partial \chi_g ^-)(\partial \chi_g ^+) + (\partial \chi^{--}_g)(\partial \chi_g^{++}) + \nonumber \\
& + \frac{g^2}{4} \chi_u^0 \chi_u^0 V^- V^+ + \frac{g^2}{4} \chi_u^0 \chi_u^0 U^{--} U^{++} + \nonumber \\
& + \frac{n_{33}^2}{2} \chi_u^0 \chi_u^0 Z' Z' + n_{33} (\partial \chi_g ^0) \chi_u^0 Z' \nonumber \\
& + i \frac{g}{2} (\partial \chi_g ^+) \chi_u^0 V^- - i \frac{g}{2} (\partial \chi_g ^-) \chi_u^0 V^+ \nonumber \\ 
& + i \frac{g}{2} (\partial \chi_g ^{++}) \chi_u^0 U^{--} - i \frac{g}{2} (\partial \chi_g ^{--}) \chi_u^0 U^{++} + \nonumber \\ 
& + \mathcal{O}(\text{non-quadratic terms})  \qquad (\beta = - \sqrt{3})
\end{eqnarray}

We remember that the gauge-fixing factor $\xi_{GB}$ may be defined independently for each gauge-boson and will consider our whole analysis in the arbitrary t'Hooft gauge. The total Lagrangian to the scalars is given by 
\begin{equation}\label{Eq:SLag}
\mathcal{L}_s = (D_\mu \chi)^\dagger(D^\mu \chi) + (D_\mu \eta)^\dagger(D^\mu \eta) + (D_\mu \rho)^\dagger(D^\mu \rho)
\end{equation}
and accounts for the gauge-bosons acquisition of mass. At this point we find one of the possible arguments to explain the absence of connection with the SM through non-diagonal interactions. First, we designate the pairs $12, 13, 23$ of the triplet entries in Eq.(\ref{Eq:Scadef}) as the first, second and third doublets, respectively. Now, since the generators $\mathbb{I}_\pm$ will act only in the first doublet, the above Lagrangian always present the vertices with the 3-component in pairs of exotic particles, forbidding all the tree-level contributions. 

From Eq.(\ref{Eq:selfintchi}) we can extract the mixed quadratic term
\begin{equation}\label{Eq:ScaV1}
\mathcal{L}_s \supset u \frac{g}{2} (\partial \chi_V ^+) V^- - u \frac{g}{2} (\partial \chi_V ^-) V^+
\end{equation}
These interactions, in general, are related to a transversal propagator for the gauge-bosons and can be equivalently replaced via the insertion of a gauge-fixing Lagrangian. Apart from that, the scalars involved in the second breaking will also contribute to the mass of $V$ and $U$, including a similar momentum-dependent vertex
\begin{equation}\label{Eq:ScaV2}
\mathcal{L}_s \supset v_\eta \frac{g}{2} (\partial \eta_V^-) V^+ - v_\eta \frac{g}{2} (\partial \eta_V^+) V^-
\end{equation}

The Goldstones will be identified after the rotation of the correlated degrees of freedom inside $\chi$ and $\eta$. To the above case, it can be achieved via the orthogonal matrix
\begin{equation}\label{Eq:Retachi}
R^V_{\eta \chi} = \frac{1}{\overline{u}_\eta} \begin{pmatrix}
u & -v_\eta \\ v_\eta & u
\end{pmatrix}, \qquad \overline{u}_\eta = \sqrt{u^2 + v_\eta^2} 
\end{equation}
or
\begin{equation}\label{Eq:roteta}
\begin{pmatrix}
u, & -v_\eta
\end{pmatrix} \partial 
\begin{pmatrix}
\chi_V \\ \eta_V
\end{pmatrix} V^- \rightarrow
\begin{pmatrix}
\overline{u}_\eta, & 0
\end{pmatrix} \partial 
\begin{pmatrix}
\overline{\chi}_V \\ \overline{\eta}_V
\end{pmatrix} V^-
\end{equation}
where
\begin{equation}\label{Eq:roteta2}
\begin{pmatrix}
\overline{\chi}_V \\ \overline{\eta}_V
\end{pmatrix} = R^V_{\eta \chi} \begin{pmatrix}
\chi_V \\ \eta_V
\end{pmatrix} 
\end{equation}
From Eq.(\ref{Eq:roteta}) we see that $\overline{\chi}_V$ can be identified as the Goldstone of our theory. The procedure thus described follows from the group structure and is independent of $\beta$. In Section \ref{Sec:A7} we will see that the rotation Eq.(\ref{Eq:roteta2}) defines a larger block diagonal matrix that plays an important role on the diagonalization of the total mass matrix. After the insertion of additional self-interactions, the potential acquires a new structure of mixing that differs from the gauge-fixing mass matrix. However, we may consistently make use of a general consequence of the Goldstone theorem which constrains these two components to lie on orthogonal subspaces, and then verify that our previous knowledge about the rotation of $\mathbb{M}_\xi$ can simplify the procedure of finding the total mass matrix, $\mathbb{M}_s$, diagonal.

In the context of Eq.(\ref{pot2}), the new matrix on the basis $\left(\chi_V \ \eta_V\right)$ will appear like
\begin{equation}
\begin{pmatrix}
\mu_\chi + \lambda_\chi u^2 + \frac{v_\eta ^2}{2}\overline{\lambda}_{\eta \chi} + \frac{v_\eta ^2}{2}\lambda_{\eta \chi} + \frac{v_\rho ^2}{2}\lambda_{\rho \chi} 
& \frac{u v_\eta}{2}\overline{\lambda}_{\eta \chi} + v_\rho \zeta
\\ \frac{u v_\eta}{2}\overline{\lambda}_{\eta \chi} + v_\rho \zeta 
& \mu_\eta + \lambda_\eta v_\eta^2 + \frac{u^2}{2}\overline{\lambda}_{\eta \chi} + \frac{u^2}{2}\lambda_{\eta \chi} + \frac{v_\rho ^2}{2}\lambda_{\rho \eta}
\end{pmatrix}
\end{equation}
which can be simplified after we apply the VSC in Eq.(\ref{Eq:vsto}) to
\begin{eqnarray}\label{Eq:meta}
V(\chi, \rho, \eta) \supset & \begin{pmatrix}
\chi^*_V & \eta^*_V
\end{pmatrix}
\begin{pmatrix}
\frac{v_\eta ^2}{2}\overline{\lambda}_{\eta \chi} + \frac{v_\rho v_\eta}{u} \zeta
& \frac{u v_\eta}{2}\overline{\lambda}_{\eta \chi} + v_\rho \zeta
\\ \frac{u v_\eta}{2}\overline{\lambda}_{\eta \chi} + v_\rho \zeta 
& \frac{u^2}{2}\overline{\lambda}_{\eta \chi} + \frac{u v_\rho }{v_\eta}\zeta
\end{pmatrix}
\begin{pmatrix}
\chi_V \\ \eta_V
\end{pmatrix} \nonumber \\ 
& = \frac{\lambda_V}{2}\begin{pmatrix}
\chi^*_V & \eta^*_V
\end{pmatrix}
\begin{pmatrix}
v_\eta ^2
& u v_\eta
\\ u v_\eta
& u^2
\end{pmatrix}
\begin{pmatrix}
\chi_V \\ \eta_V
\end{pmatrix}
\end{eqnarray}
where 
\begin{equation}
\frac{\lambda_V}{2} = \frac{\overline{\lambda}_{\eta \chi}}{2} +\frac{ v_\rho \zeta}{u v_\eta}
\end{equation}
The matrix in Eq.(\ref{Eq:meta}) can also be directly diagonalized via $R^V_{\eta \chi}$, leading to
\begin{equation}\label{Eq:etamass}
\frac{\lambda_V}{2}
\begin{pmatrix}
\overline{\chi}^*_V & \overline{\eta}^*_V
\end{pmatrix}
\begin{pmatrix}
0 & 0 \\ 0 & \overline{u}_\eta^2
\end{pmatrix}
\begin{pmatrix}
\overline{\chi}_V \\ \overline{\eta}_V
\end{pmatrix} \quad \rightarrow \quad m^2_\eta = \lambda_V \overline{u}_\eta^2
\end{equation}

We are then verifying the Goldstone theorem being manifest. The quadratic mass coming from the gauge-fixing Lagrangian is
\begin{equation}\label{Eq:GFV}
\mathbb{M}^2_\xi = \xi_V \begin{pmatrix}
u^2 & -u v_\eta \\ -u v_\eta & v_\eta ^2
\end{pmatrix} \quad \rightarrow \quad \mathbb{M}^2_\xi \cdot \begin{pmatrix}
v_\eta ^2
& u v_\eta
\\ u v_\eta
& u^2
\end{pmatrix} = \mathbb{O}
\end{equation}
and we have registered its orthogonality to the matrix in Eq.(\ref{Eq:meta}).

\subsection{Gauge-Boson Masses}\label{Sec:A5}
The gauge-dependent mass matrix is sufficient to exhibit the gauge-boson masses and, for completeness, we present them here as extracted from the Scalar Lagrangian:
\begin{equation}
m^2_W = \frac{g^2}{4} \overline{v}^2, \qquad m^2_V = \frac{g^2}{4} \overline{u}_\rho^2, \qquad m^2_U = \frac{g^2}{4} \overline{u}_\eta^2, \qquad
m^2_Z = \frac{g^2}{4 c_w^2}  \overline{v}^2
\end{equation}

In order to be general we distinguish the $U$ and $V$ masses above. However, on the assumption $u \gg v_\rho, v_\eta$, their values would be equal and different of the $Z'$ via a $\beta$ dependent factor. For $\beta = - \sqrt{3}$ we would find
\begin{equation}
m^2_{Z'} = \frac{g^2}{3}  u^2  c_x^2 \qquad (\beta = - \sqrt{3})
\end{equation}
The bilinears for neutral vectors are given by
\begin{equation}
\mathcal{L}_s \supset -g \frac{c_x}{\sqrt{3}} \partial \chi_g Z' - g \frac{v_\rho}{2c_w} \partial \rho_g Z + g \frac{v_\eta}{2c_w} \partial \eta_g Z 
\end{equation}
corresponding, after the $\mathcal{L}_{g.f.}$ insertion, to the correct $\xi$-dependent $Z,Z'$ propagators and a new mass matrix to the scalars:
\begin{equation}
\mathbb{M}^2_{\xi}\Bigl|_{Z,Z'} = 
\begin{pmatrix}
\frac{\xi_Z}{4 c_w^2}
\begin{pmatrix}
v_\eta^2 & -v_\eta v_\rho \\ -v_\eta v_\rho & v_\rho^2
\end{pmatrix} & \mathbf{0} \\
\mathbf{0}^\intercal & \xi_{Z'}\frac{u^2}{3} c_x^2
\end{pmatrix}
\end{equation}
on the basis $\left(\eta_g, \rho_g, \chi_g \right)$. It can be verified that the above result is in fact orthogonal to the $\mathbb{P}$ matrix presented in the next section.

\subsection{Scalar Masses}\label{Sec:A6}

We introduce this section approaching the masses for the remaining scalars. The expansion of Eq.(\ref{pot2}) give us the pattern of mixing between the charged particles and a result similar to Eq.(\ref{Eq:etamass}) must arise. Let us cover this explicitly, but now in a more direct mode. 

For the U-type scalars on the basis $\left(\chi_U \ \rho_U\right)$ we have
\begin{equation}
\begin{pmatrix}
\mu_\chi + \lambda_\chi u^2 + \frac{v_\rho ^2}{2}\overline{\lambda}_{\rho \chi} + \frac{v_\eta ^2}{2}\lambda_{\eta \chi} + \frac{v_\rho ^2}{2}\lambda_{\rho \chi} 
& \frac{u v_\eta}{2}\overline{\lambda}_{\rho \chi} + v_\eta \zeta
\\ \frac{u v_\eta}{2}\overline{\lambda}_{\rho \chi} + v_\eta \zeta 
& \mu_\rho + \lambda_\rho v^2 + \frac{u^2}{2}\overline{\lambda}_{\rho \chi} + \frac{u^2}{2}\lambda_{\rho \chi} + \frac{v_\eta ^2}{2}\lambda_{\rho \eta}
\end{pmatrix}
\end{equation}
With the help of Eq.(\ref{Eq:vsto}) it can be simplified to
\begin{equation}
\frac{\lambda_U}{2}\begin{pmatrix}
\chi^*_U & \rho^*_U
\end{pmatrix}
\begin{pmatrix}
v_\rho ^2
& u v_\rho
\\ u v_\rho
& u^2
\end{pmatrix}
\begin{pmatrix}
\chi_U \\ \rho_U
\end{pmatrix} \quad \rightarrow \quad
\frac{\lambda_U}{2} = \frac{\overline{\lambda}_{\rho \chi}}{2} +\frac{ v_\eta \zeta}{u v_\rho}
\end{equation}
or
\begin{equation}\label{Eq:rhomass}
\frac{\lambda_U}{2}
\begin{pmatrix}
\overline{\chi}^*_U & \overline{\rho}^*_U
\end{pmatrix}
\begin{pmatrix}
0 & 0 \\ 0 & \overline{u}_\rho^2
\end{pmatrix}
\begin{pmatrix}
\overline{\chi}_U \\ \overline{\rho}_U
\end{pmatrix} \quad \rightarrow \quad m^2_\rho = \lambda_U \overline{u}_\rho^2
\end{equation}
where the rotation matrix is given by
\begin{equation}
R^U_{\rho \chi} = \frac{1}{\overline{u}_\rho} \begin{pmatrix}
u & -v_\rho \\ v_\rho & u
\end{pmatrix} \quad \rightarrow \quad \overline{u}_\rho = \sqrt{u^2 + v_\rho^2} 
\end{equation}

To conclude the non-diagonal sector, there are those scalars coupled with the SM gauge bosons:
\begin{equation}
\frac{\lambda_W}{2}\begin{pmatrix}
\eta^*_W & \rho^*_W
\end{pmatrix}
\begin{pmatrix}
v_\rho ^2
& v_\rho v_\eta
\\ v_\rho v_\eta
& v_\eta ^2
\end{pmatrix}
\begin{pmatrix}
\eta_W \\ \rho_W
\end{pmatrix} \quad \rightarrow \quad
\frac{\lambda_W}{2} = \frac{\overline{\lambda}_{\rho \eta}}{2} +\frac{u \zeta}{v_\rho v_\eta}
\end{equation}
or
\begin{equation}\label{Eq:rhosm}
\frac{\lambda_W}{2}
\begin{pmatrix}
\overline{\eta}^*_W & \overline{\rho}^*_W
\end{pmatrix}
\begin{pmatrix}
0 & 0 \\ 0 & \overline{v}^2
\end{pmatrix}
\begin{pmatrix}
\overline{\eta}_W \\ \overline{\rho}_W
\end{pmatrix} \quad \rightarrow \quad m^2_{\rho_W} = \lambda_W \overline{v}^2
\end{equation}
Equivalently,
\begin{equation}
R^W_{\rho \eta} = \frac{1}{\overline{v}} \begin{pmatrix}
v_\eta & - v_\rho \\ v_\rho & v_\eta
\end{pmatrix} \quad \rightarrow \quad \overline{v} = \sqrt{v_\rho^2 + v_\eta^2} 
\end{equation}
where
\begin{equation}\label{Eq:rhopmass}
\begin{pmatrix}
\overline{\eta}_W \\ \overline{\rho}_W
\end{pmatrix} = R^W_{\rho \eta} \begin{pmatrix}
\eta_W \\ \rho_W
\end{pmatrix} 
\end{equation}

From the beginning of our review we have assumed $Z'$ to not mix with $Z$ and Photon, and justified this claim because of the hierarchy between the breaking scales. We may expect, therefore, the same pattern to be also present in the neutral scalar sector. On what follows we try to identify this feature in detail.   

First we write the mass matrix of pseudo-scalars on the following basis
\begin{equation}
\begin{pmatrix}
\eta_g & \rho_g & \chi_g
\end{pmatrix} \mathbb{P}
\begin{pmatrix}
\eta_g \\ \rho_g \\ \chi_g
\end{pmatrix}
\end{equation}
such that the entries of $\mathbb{P}$ are given by
\begin{equation}
\begin{pmatrix}
\frac{1}{2}(\mu_\eta + \lambda_\eta v_\eta^2 + \frac{u^2}{2}\lambda_{\eta \chi} + \frac{v_\rho ^2}{2}\lambda_{\rho \eta}) & \frac{u \zeta}{2} & \frac{v_\rho \zeta}{2} \\
\frac{u \zeta}{2} & \frac{1}{2}(\mu_\rho + \lambda_\rho v_\rho ^2 + \frac{u^2}{2} \lambda_{\rho \chi} + \frac{v_\eta ^2}{2} \lambda_{\rho \eta}) & \frac{v_\eta \zeta}{2} \\
\frac{v_\rho \zeta}{2} & \frac{v_\eta \zeta}{2} & \frac{1}{2}(\mu_\chi + \lambda_\chi u^2 + \frac{v_\eta ^2}{2}\lambda_{\eta \chi} + \frac{v_\rho ^2}{2} \lambda_{\rho \chi})
\end{pmatrix}	
\end{equation}
or, simply, 
\begin{equation}
\mathbb{P} = \frac{u \zeta}{2}
\begin{pmatrix}
\frac{v_\rho }{v_\eta}& 1 & \frac{v_\rho}{u} \\
1 & \frac{v_\eta }{v_\rho} & \frac{v_\eta}{u} \\
\frac{v_\rho}{u} & \frac{v_\eta}{u} & \frac{v_\rho v_\eta}{u^2}
\end{pmatrix} \approx \frac{u \zeta}{2}
\begin{pmatrix}
\frac{v_\rho}{v_\eta}& 1 & 0 \\
1 & \frac{v_\eta}{v_\rho} & 0 \\
0 & 0 & 0
\end{pmatrix} = 
\frac{\lambda_P}{2}
\begin{pmatrix}
v_\rho ^2& v_\rho v_\eta & 0 \\
v_\rho v_\eta & v_\eta ^2 & 0 \\
0 & 0 & 0
\end{pmatrix}	
\end{equation}
We emphasize that the second signal is representing the absence of contribution from the low-energy breaking to the $Z'$ mass and must not be understood as a formal limit. The constant is given by
\begin{equation}
\lambda_P = \frac{u \zeta}{v_\rho v_\eta}
\end{equation}
Thus the pseudo-scalars are diagonalized to their mass eigenstates by the same matrix $R^W_{\rho \eta}$ that the standard charged scalars. Finally, the potential introduces a single pseudo-scalar, that we called $\rho_P$, such that
\begin{equation}
\rho_P: \qquad m^2_{P} = \lambda_P \overline{v}^2 
\end{equation}
The presence of this particle is $\beta$-independent.

The real scalars follow a similar analysis. In the basis
\begin{equation}
\begin{pmatrix}
H_\eta & H_\rho & \overline{H}
\end{pmatrix} \mathbb{S}
\begin{pmatrix}
H_\eta \\ H_\rho \\ \overline{H}
\end{pmatrix}
\end{equation}
and after the Eq.(\ref{Eq:vsto}) insertion, the $\mathbb{S}$ entries are
\begin{equation}
\mathbb{S} = \begin{pmatrix}
\lambda_\eta v_\eta^2  + \zeta \frac{u v_\rho}{2 v_\eta} & \lambda_{\rho \eta} \frac{v_\eta v_\rho}{2} - \frac{u \zeta}{2} & 0 \\
\lambda_{\rho \eta} \frac{v_\eta v_\rho}{2} - \frac{u \zeta}{2} & \lambda_\rho v_\rho ^2  + \zeta \frac{u v_\eta}{2 v_\rho} & 0 \\
0 & 0 & \lambda_\chi u^2  + \zeta \frac{v_\eta v_\rho}{2 u}
\end{pmatrix}	
\end{equation}
The above result includes that, in the context of $u \gg v_\rho v_\eta$, the third non-diagonal elements are small compared with the remaining and thus give a neglected contribution to the eigenvalues. We register their original values:
\begin{eqnarray}
\mathbb{S}_{13} &:& \zeta \frac{v_\rho}{2} - \frac{v_\eta}{u} (\mu_\eta + \lambda_\eta v_\eta^2) - \lambda_{\rho \eta} \frac{v_\rho^2 v_\eta}{2 u} \\
\mathbb{S}_{23} &:& \zeta \frac{v_\eta}{2} - \frac{v_\rho}{u} (\mu_\rho + \lambda_\rho v_\rho^2) - \lambda_{\rho \eta} \frac{v_\eta^2 v_\rho}{2 u} 
\end{eqnarray}
The matrix $\mathbb{S}$ can be diagonalized by
\begin{equation}\label{Eq:RS}
R_S = \begin{pmatrix}
c_s & -s_s & 0 \\
s_s & c_s & 0 \\
0 & 0 & 1
\end{pmatrix}
\end{equation}
and their simplified eigenvalues are:
\begin{eqnarray}
m^2_{H_\eta} &=& \frac{1}{v_\eta v_\rho}(\lambda_\eta v_\eta^3 v_\rho + \lambda_\rho v_\rho^3 v_\eta) \nonumber \\
&=& \lambda_\eta v_\eta^2 + \lambda_\rho v_\rho^2 \\
m^2_{H_\rho} &=& \frac{1}{v_\eta v_\rho}(\lambda_\eta v_\eta^3 v_\rho + \lambda_\rho v_\rho^3 v_\eta + u \zeta( v_\rho^2 + v_\eta^2)) \nonumber \\
& = & \lambda_\eta v_\eta^2 + \lambda_\rho v_\rho^2 + u \zeta\frac{(v_\eta^2 + v_\rho^2)}{v_\eta v_\rho} \\
m^2_{\overline{H}} &=& 2 \lambda_\chi u^2 + \zeta \frac{v_\eta v_\rho}{u}
\end{eqnarray}
Note, therefore, an approximate relation between the scalar and pseudo-scalar masses:
\begin{equation}
m^2_{H_\rho} = m^2_{H_\eta} + m^2_P
\end{equation}

Thus, as in the Standard Model, the Vacuum Stability Condition prevents the mass terms of some charged scalars to arise in the potential, leaving their contribution exclusively in the gauge-fixing Lagrangian. 
In the previous discussion we identified this feature occurring for our version-independent 3-3-1HL potential and the VSC leading us to a correct scalar spectrum. Nevertheless, we have not mentioned yet about the allowed $\beta$-dependent pieces, like for example
\begin{equation}\label{Eq:goldstone}
V(\chi, \rho, \eta)\Bigr|_{\beta = - \sqrt{3}} \ \supset \ V^\chi_{\rho \eta} \equiv \lambda^\chi _{\rho \eta}(\chi^\dagger \eta)(\rho^\dagger \eta) + h.c.
\end{equation}
which could connect the scalars $\left(\eta_W, \rho_W\right)$ with $\left(\chi_V, \eta_V\right)$. Along the rest of this work we verify that the omission of a term like Eq.(\ref{Eq:goldstone}) may imply a complete dissociation of the charged exotic sector with standard particles at tree-level, leaving the Yukawa interactions as the ultimate chance to connect them. In Section (\ref{Sec:A8}) we disclose this possibility and conclude that assuming Eq.(\ref{Eq:goldstone}) equal to zero is equivalent to assume a discrete symmetry for this variant of the 3-3-1HL.

\subsection{The Potential for particular models}\label{Sec:A7}

As mentioned previously, the term $\lambda^\chi _{\rho \eta}(\chi^\dagger \eta)(\rho^\dagger \eta)$ could a priori be included in the potential\footnote{For $\beta = + \sqrt{3}$ the new term is similar to Eq.(\ref{Eq:goldstone}), with $\rho \leftrightarrow \eta$.} for $\beta = - \sqrt{3}$, creating a leading order connection with the SM from one additional mixing between the four charged scalars.

First, let us see how the mass matrix will emerge in the basis $\begin{pmatrix}
\eta_W & \rho_W & \eta_V & \chi_V
\end{pmatrix}$
after this insertion:
\begin{equation}\label{Eq.:PotMatrix}
\mathbb{M}^2_V = 
\begin{pmatrix}
\frac{\lambda_W}{2} \left(\begin{matrix}
v_\rho ^2
& v_\rho v_\eta
\\ v_\rho v_\eta
& v_\eta ^2
\end{matrix}\right)
& \lambda^\chi _{\rho \eta} \left(\begin{matrix}
u v_\rho
& v_\eta v_\rho
\\ u v_\eta
& v_\eta^2
\end{matrix}\right)
\\
\lambda^\chi _{\rho \eta} \left(\begin{matrix}
u v_\rho
& u v_\eta
\\ v_\eta v_\rho
& v_\eta^2
\end{matrix}\right)
& \frac{\lambda_V}{2}\left(\begin{matrix}
u ^2
& u v_\eta
\\ u v_\eta
& v_\eta ^2
\end{matrix}\right)
\end{pmatrix}
\end{equation} 
We observe that the determinant of $\mathbb{M}^2_V$ above is in fact zero, a necessary condition to leave the theory with a well-defined spectrum.

The gauge-fixing Lagrangian also contributes with a gauge-dependent squared mass-matrix, and we have:
\begin{equation}\label{Eq:Mxi}
\mathbb{M}^2_\xi = 
\begin{pmatrix}
\xi_W \begin{pmatrix}
v_\eta ^2
& - v_\rho v_\eta
\\ - v_\rho v_\eta
& v_\rho ^2
\end{pmatrix}
& \mathbb{O}
\\
\mathbb{O}
& \xi_V \begin{pmatrix}
v_\eta ^2 
& - u v_\eta
\\ - u v_\eta
& u ^2
\end{pmatrix}
\end{pmatrix}
\end{equation} 

As mentioned before, one important consequence of Goldstone theorem requires that these two matrices must lie in orthogonal subspaces, such that
\begin{equation}\label{Eq:ortho}
\mathbb{M}^2_V \cdot \mathbb{M}^2_\xi = 0  \qquad (\mathbb{M}^2_\xi \cdot \mathbb{M}^2_V = 0) 
\end{equation}
We can check this directly if we remember a simple identity for matrix products. Consider a general even $n \times n$ matrix:
\begin{equation}
\mathbb{M} = \begin{pmatrix}
\mathbb{M}_1 & \mathbb{M}_2 \\
\mathbb{M}_3 & \mathbb{M}_4 
\end{pmatrix}
\end{equation}
where $\mathbb{M}_k$ are $\frac{n}{2} \times \frac{n}{2}$. Then, it follows that
\begin{equation}\label{Eq:matrixrule}
\mathbb{A}\cdot\mathbb{B} = \begin{pmatrix}
\left(\mathbb{A}_1 \cdot \mathbb{B}_1 + \mathbb{A}_2 \cdot \mathbb{B}_3\right) & \left(\mathbb{A}_1 \cdot \mathbb{B}_2 + \mathbb{A}_2 \cdot \mathbb{B}_4\right) \\
\left(\mathbb{A}_3 \cdot \mathbb{B}_1 + \mathbb{A}_4 \cdot \mathbb{B}_3\right) & \left(\mathbb{A}_3 \cdot \mathbb{B}_2 + \mathbb{A}_4 \cdot \mathbb{B}_4\right)
\end{pmatrix}
\end{equation}
which can be proved by considering a single element
\begin{equation}
\mathbb{M}_{ij} \in \begin{cases}
\mathbb{M}_{1}, \quad  i \in [1, \dots \frac{n}{2}], j \in  [1, \dots \frac{n}{2}] \\
\mathbb{M}_{2}, \quad  i \in [1, \dots \frac{n}{2}], j \in  [\frac{n}{2} + 1, \dots n] \\
\mathbb{M}_{3}, \quad  i \in [\frac{n}{2} + 1, \dots n], j \in [1, \dots \frac{n}{2}] \\
\mathbb{M}_{4}, \quad  i \in [\frac{n}{2} + 1, \dots n], j \in [\frac{n}{2} + 1, \dots n]
\end{cases}
\end{equation}
and the general rule for matrix products, $(\mathbb{A}\cdot\mathbb{B})_{ij} = \mathbb{A}_{ik}\mathbb{B}_{kj} + \mathbb{A}_{il}\mathbb{B}_{lj}$, with $k \in [1, \dots \frac{n}{2}]$ and $l \in  [\frac{n}{2} + 1, \dots n]$. For instance, if $(ij) \in \mathbb{M}_2$ then $(\mathbb{A}\cdot\mathbb{B})_{ij} = (\mathbb{A}_1 \cdot \mathbb{B}_2 + \mathbb{A}_2 \cdot \mathbb{B}_4)_{i(j-\frac{n}{2})}$.

Finally, the Eq.(\ref{Eq:matrixrule}), for $\mathbb{B}_2 = \mathbb{B}_3 = 0$, implies
\begin{equation}\label{Eq:prodident}
\mathbb{A}\cdot\mathbb{B} = \begin{pmatrix}
\left(\mathbb{A}_1 \cdot \mathbb{B}_1\right) & \left(\mathbb{A}_2 \cdot \mathbb{B}_4\right) \\
\left(\mathbb{A}_3 \cdot \mathbb{B}_1\right) & \left(\mathbb{A}_4 \cdot \mathbb{B}_4\right)
\end{pmatrix}
\end{equation}
and the demonstration of Eq.(\ref{Eq:ortho}) follows straightforwardly.
\begin{itemize}
	\item The result of Eq.(\ref{Eq:matrixrule}) can be also extended for odd matrices, and to the $3 \times 3$ case we register one useful identity. First, consider the generic matrix 
	\begin{equation}
	\mathbb{M} = \begin{pmatrix}
	\mathbb{M}_1 & \mathbf{m}_2 \\
	\mathbf{m}^\intercal_3 & m_4 
	\end{pmatrix}
	\end{equation}
	where $\mathbf{m}$ are two-dimensional vectors and $m_4$ is a c-number. Thus, we have
	\begin{equation}\label{Eq:matrix3rule}
	\mathbb{A}\cdot\mathbb{B} = \begin{pmatrix}
	\left(\mathbb{A}_1 \cdot \mathbb{B}_1 + \mathbf{a}_2 \cdot \mathbf{b}^\intercal_3\right) & \left(\mathbb{A}_1 \cdot \mathbf{b}_2 + \mathbf{a}_2 b_4\right) \\
	\left(\mathbf{a}^\intercal_3 \cdot \mathbb{B}_1 + a_4 \mathbf{b}^\intercal_3\right) & \mathbf{a}^\intercal_3 \cdot \mathbf{b}_2 + a_4 b_4
	\end{pmatrix}
	\end{equation}	
\end{itemize}

Now, let us define the total squared mass
\begin{equation}\label{Eq:totmass}
\mathbb{M}^2_s \equiv \mathbb{M}^2_V + \mathbb{M}^2_\xi  
\end{equation}
From Eq.(\ref{Eq:ortho}) it follows that
\begin{equation}
\mathbb{M}^2_s \cdot \mathbb{M}^2_\xi  = \mathbb{M}^2_\xi \cdot \mathbb{M}^2_s =(\mathbb{M}^2_\xi)^2 
\end{equation}
i.e. these matrices must commute:
\begin{equation}\label{Eq:comut}
\left[\mathbb{M}^2_s,\mathbb{M}^2_\xi \right] = 0
\end{equation}
and, therefore, there must be a matrix $\mathbb{U}$ able to diagonalize them simultaneously.
Apart from that, we have already seen that the matrix
\begin{equation}\label{Eq:diag}
\mathbb{D} = \begin{pmatrix}
\frac{1}{\overline{v}}\begin{pmatrix}
v_\eta
& - v_\rho
\\ v_\rho
& v_\eta
\end{pmatrix}
& \mathbb{O}
\\
\mathbb{O}
& \frac{1}{\overline{u}_\eta}\begin{pmatrix}
u 
& v_\eta
\\ -v_\eta
& u
\end{pmatrix}
\end{pmatrix}
\end{equation}
is such that 
\begin{equation}
\mathbb{D} \mathbb{M}^2_\xi \mathbb{D}^\intercal = \mathbb{X}^2_\xi
\end{equation}
with $\mathbb{X}^2_\xi$ diagonal. Thus, from Eq.(\ref{Eq:comut}), it follows that $\mathbb{D}\left[\mathbb{M}^2_s,\mathbb{M}^2_\xi \right] \mathbb{D}^\intercal = 0 $, or
\begin{equation}\label{Eq.:Comute}
\left[\overline{\mathbb{M}}^2_s,\mathbb{X}^2_\xi\right] = 0, \quad \textrm{where} \quad \overline{\mathbb{M}}^2_s \equiv \mathbb{D}\mathbb{M}^2_s\mathbb{D}^\intercal
\end{equation}
i.e. the rotation of the total mass matrix by $\mathbb{D}$ will commute with the diagonal mass matrix of the Goldstone bosons. In terms of components the above result can be written as
\begin{equation}
\sum_k (\overline{\mathbb{M}}^2_s)_{ik} (\mathbb{X}^2_\xi)_{kj} = \sum_k (\mathbb{X}^2_\xi)_{ik}(\overline{\mathbb{M}}^2_s)_{kj}
\end{equation} 
or
\begin{equation}
\sum_k \left[(\overline{\mathbb{M}}^2_s)_{ik} \delta_{kj}(\mathbb{X}^2_\xi)_{jj} - \delta_{ik}(\overline{\mathbb{M}}^2_s)_{kj}(\mathbb{X}^2_\xi)_{ii}\right] = 0
\end{equation} 
and, finally,
\begin{equation}\label{Eq.:Block}
(\overline{\mathbb{M}}^2_s)_{ij}\left\{(\mathbb{X}^2_\xi)_{ii} - (\mathbb{X}^2_\xi)_{jj}\right\} = 0
\end{equation} 
Thus, the $\overline{\mathbb{M}}^2_s$ must emerge into a block diagonal form, whose dimension will agree with the number of eigenvalues inside $\mathbb{X}^2_\xi$ with multiplicity one.

Our second step consists on finding  a matrix $\overline{\mathbb{D}}$ to diagonalize $\overline{\mathbb{M}}^2_s$ and, in general, simple to define. Again, from Eq.(\ref{Eq.:Comute}), the matrix $\mathbb{X}^2_\xi$ would remain block-diagonal after a rotation by $\overline{\mathbb{D}}$ and, in our case we find
\begin{equation}
(\overline{\mathbb{D}}\mathbb{D})\mathbb{M}^2_\xi(\overline{\mathbb{D}}\mathbb{D})^\intercal = \mathbb{X}^2_\xi \quad \textrm{and} \quad (\overline{\mathbb{D}}\mathbb{D})\mathbb{M}^2_s(\overline{\mathbb{D}}\mathbb{D})^\intercal = \mathbb{X}^2_s
\end{equation}
with $\mathbb{X}^2_s$ diagonal.

In other words, we are finding by parts the matrix that can simultaneously diagonalize the total and the gauge-fixing mass matrices, after claiming a general Goldstone theorem corollary. Surely, we just need to look at the total $\mathbb{M}^2_s$. However, as we can dissociate it from the gauge-dependent part, it becomes clear through these steps the direct identification of the Goldstone bosons.

Let us apply these results in our particular case, from Eq.(\ref{Eq.:PotMatrix}) and Eq.(\ref{Eq:Mxi}). As we previously obtained, $\mathbb{X}^2_\xi$ is given by
\begin{equation}
\mathbb{X}^2_\xi = 
\begin{pmatrix}
\xi_W \begin{pmatrix}
\overline{v}^2
& 
\\ 
& 0
\end{pmatrix}
& \mathbb{O}
\\
\mathbb{O}
& \xi_V \begin{pmatrix}
0
& 
\\ 
& \overline{u}_\eta^2
\end{pmatrix}
\end{pmatrix}
\end{equation} 
and, from Eq.(\ref{Eq.:Block}), we must expect $(\mathbb{D})\mathbb{M}^2_s(\mathbb{D})^\intercal$ to contain a $2 \times 2$ block. In fact,
\begin{equation}
(\mathbb{D})\mathbb{M}^2_s(\mathbb{D})^\intercal = \begin{pmatrix}
\xi_W \overline{v}^2 & & \\ 
 & \begin{pmatrix}
\frac{\lambda_W}{2} \overline{v}^2 & \lambda^\chi _{\rho \eta} \ \overline{v} \ \overline{u}_\eta \\ \lambda^\chi _{\rho \eta} \ \overline{v} \ \overline{u}_\eta & \frac{\lambda_V}{2} \overline{u}_\eta^2
 \end{pmatrix} & \\
 & & \xi_V \overline{u}_\eta^2
\end{pmatrix}
\end{equation}
The central matrix can be diagonalized after the insertion of a new $\overline{\mathbb{D}}$, such that
\begin{equation}
\overline{\mathbb{D}} =  \begin{pmatrix}
1 & & \\ 
& \frac{1}{\overline{y}}\begin{pmatrix}
a_1 & \lambda^\chi _{\rho \eta} \ \overline{v} \ \overline{u}_\eta \\ \lambda^\chi _{\rho \eta} \ \overline{v} \ \overline{u}_\eta & a_4
\end{pmatrix} & \\
& & 1
\end{pmatrix}
\end{equation}

\begin{eqnarray}
a_1 &\equiv& \frac{1}{2}\left(\frac{\lambda_W}{2} \overline{v}^2 - \frac{\lambda_V}{2} \overline{u}_\eta^2 - \sqrt{\left(\frac{\lambda_W}{2} \overline{v}^2 - \frac{\lambda_V}{2} \overline{u}_\eta^2\right)^2 + (2\lambda^\chi _{\rho \eta} \ \overline{v} \ \overline{u}_\eta)^2}\right) \\
a_4 &\equiv& \frac{1}{2}\left( \frac{\lambda_V}{2} \overline{u}_\eta^2 -\frac{\lambda_W}{2} \overline{v}^2 + \sqrt{\left(\frac{\lambda_W}{2} \overline{v}^2 - \frac{\lambda_V}{2} \overline{u}_\eta^2\right)^2 + (2\lambda^\chi _{\rho \eta} \ \overline{v} \ \overline{u}_\eta)^2}\right)\\
\overline{y} &\equiv & \sqrt{(\lambda^\chi _{\rho \eta} \ \overline{v} \ \overline{u}_\eta)^2 + a_1^2}
\end{eqnarray}
and, finally,
\begin{equation}
\mathbb{X}^2_s \equiv (\mathbb{D}\overline{\mathbb{D}})\mathbb{M}^2_s(\mathbb{D}\overline{\mathbb{D}})^\intercal = \begin{pmatrix}
\xi_W \overline{v}^2 & & \\ 
& \begin{pmatrix}
m^2_{\rho_W} & \\ 
 & m^2_{\eta}
\end{pmatrix} & \\
& & \xi_V \overline{u}_\eta^2
\end{pmatrix}
\end{equation}
where
\begin{eqnarray}
m^2_{\rho_W} &\equiv& \frac{1}{2}\left(\frac{\lambda_W}{2} \overline{v}^2 + \frac{\lambda_V}{2} \overline{u}_\eta^2 - \sqrt{\left(\frac{\lambda_W}{2} \overline{v}^2 - \frac{\lambda_V}{2} \overline{u}_\eta^2\right)^2 + (2\lambda^\chi _{\rho \eta} \ \overline{v} \ \overline{u}_\eta)^2}\right) \nonumber \\
m^2_{\eta} &\equiv& \frac{1}{2}\left( \frac{\lambda_W}{2} \overline{v}^2 + \frac{\lambda_V}{2} \overline{u}_\eta^2 + \sqrt{\left(\frac{\lambda_W}{2} \overline{v}^2 - \frac{\lambda_V}{2} \overline{u}_\eta^2\right)^2 + (2\lambda^\chi _{\rho \eta} \ \overline{v} \ \overline{u}_\eta)^2}\right) \nonumber
\end{eqnarray}

In summary, we have seen that the $\beta$-independent potential follows a mixing pattern in precise accordance with the $SU(2)$ subalgebras, i.e. by pairs of the scalars entailed by the same gauge-interactions. In the notation of Eq.(\ref{Eq:Scadef}) it corresponds to $\left(\chi^V \ \eta^V\right), \left(\chi^U \ \eta^U\right)$ and $\left(\rho^W \ \eta^W\right)$. When this pattern is broken by the insertion of specific allowed terms, the physical fields inside these pairs can be connected through an additional rotation, leaving the gauge-sector unchanged.

The previous analysis on the additional term in the potential holds for $|\beta| = \sqrt{3}$. In addition, we see from Eq.(\ref{Eq:Scadef}) that for $|\beta| = \frac{1}{\sqrt{3}}$ one of the triplets $\rho$ or $\eta$ has the same hypercharge as $\chi$. The authors in \cite{Sanchez-Vega:2016dwe} introduce a new non-$Z_2$ piece in the context of neutral heavy leptons, i.e for $\beta = - \frac{1}{\sqrt{3}}$, where $\chi$ and $\eta$ share the same $X$. These new contributions are given by
\begin{equation}\label{Eq:NewV}
V(\chi, \rho, \eta)\Bigr|_{\beta = - \frac{1}{\sqrt{3}}} \ \supset \ V_{\chi \eta} \equiv \mu_{\chi \eta} \chi^\dagger \eta + \lambda^\eta_{\chi \eta} (\chi^\dagger \eta) (\eta^\dagger \eta) + \lambda^\rho_{\chi \eta} (\chi^\dagger \eta) (\rho^\dagger \rho) + \lambda^\chi_{\chi \eta} (\chi^\dagger \eta) (\chi^\dagger \chi) + h.c.
\end{equation}
In this version of the model, the gauge-boson V has electric charge equals to zero. The above expression presents a linear term in the complex neutral field $\chi^0_V$, whose coefficient must be zero in order to leave $V(\chi, \rho, \eta)$ with a lower limit at $\langle \chi^0_V \rangle = 0$. This condition converts into the equation:
\begin{equation}\label{Eq:NewStab}
\mu_{\chi \eta} + \lambda^\eta_{\chi \eta} \frac{v_\eta^2}{2} + \lambda^\rho_{\chi \eta} \frac{v_\rho^2}{2} + \lambda^\chi_{\chi \eta} \frac{u^2}{2} = 0
\end{equation} 

Nevertheless, the bilinear $(\chi^\dagger \eta)$ in Eq. (\ref{Eq:NewV}) turns factorized by the r.h.s of the above expression, implying that the $\mu_{\chi \eta}$ portion along with the mixing terms for the charged scalars $\left(\chi_U \ \eta_W\right)$ will vanish. We rewrite Eq.(\ref{Eq:NewV}) as
\begin{equation}\label{Eq:NewV2}
V_{\chi \eta} = \lambda^\eta_{\chi \eta} (\chi^\dagger \eta) (\eta^\dagger \eta)\big|_{v_\eta^2 = 0} + \lambda^\rho_{\chi \eta} (\chi^\dagger \eta) (\rho^\dagger \rho)\big|_{v_\rho^2 = 0}  + \lambda^\chi_{\chi \eta} (\chi^\dagger \eta) (\chi^\dagger \chi)\big|_{u^2 = 0}  + h.c.
\end{equation}
and emphasize that the notation excludes only the quadratic v.e.v's, preserving the trilinear interactions. 

Thus far the exotic sector is still connected to the SM particles through tree-level interactions with real Higgs bosons. There will still be remaining bilinears involving the real scalars and the neutral $\left(\chi_V, \eta_V\right)$:
\begin{eqnarray}
V_{\chi \eta} &\supset& \frac{1}{\sqrt{2}}\left(\lambda^\eta_{\chi \eta} u v_\eta \ \eta_V H_\eta + \lambda^\rho_{\chi \eta} u v_\rho \ \eta_V H_\rho + \lambda^\chi_{\chi \eta} u^2 \ \eta_V \overline{H} + h.c. \right) + \nonumber \\
&+& \frac{1}{\sqrt{2}}\left(\lambda^\eta_{\chi \eta} v_\eta^2 \ \chi_V H_\eta + \lambda^\rho_{\chi \eta} v_\eta v_\rho \ \chi_V H_\rho + \lambda^\chi_{\chi \eta} u v_\eta \ \chi_V \overline{H} + h.c. \right)
\end{eqnarray}
For real $\lambda's$, these couplings mix the Higgs sector with the real part of the V-scalars, that we denote as $\left(\chi^r_V, \eta^r_V\right)$.

The gauge-fixing Lagrangian to the complex neutral boson $V$ will arise similarly to Eq.(\ref{Eq:ScaV1}) and Eq.(\ref{Eq:ScaV2}), now with complex conjugates. Thus, on the basis $\left(\chi^r_V \ \eta^r_V \ H_\eta \ H_\rho \ \overline{H} \right)$, only the first block of $\mathbb{M}_\xi$ differs from zero and is equal to the result of Eq.(\ref{Eq:GFV}):
\begin{equation}
\mathbb{M}_{\xi_V} \ = \ \begin{pmatrix}
\xi_V \begin{pmatrix}
u^2 & -u v_\eta \\ -u v_\eta & v_\eta ^2
\end{pmatrix}  & \mathbb{O} \\
\mathbb{O} & \mathbb{O}
\end{pmatrix}
\end{equation}
The contribution from $V(\chi, \rho, \eta)$ is given by
\begin{equation}
\mathbb{M}_V\Bigr|_{\beta = - \frac{1}{\sqrt{3}}} \ = \ \begin{pmatrix}
\mathbb{M}_{\chi \eta} & \mathbb{M}_2 \\
\mathbb{M}^\dagger_2 & \mathbb{S}
\end{pmatrix} \quad \text{where} \quad \mathbb{M}_2 \equiv \sqrt{2} \begin{pmatrix}
\lambda^{\eta}_{\chi \eta} u v_\eta & \lambda^{\rho}_{\chi \eta} u v_\rho & \lambda^{\chi}_{\chi \eta} u^2 \\
\lambda^{\eta}_{\chi \eta} u v_\eta & \lambda^{\rho}_{\chi \eta} u v_\rho & \lambda^{\chi}_{\chi \eta} u^2
\end{pmatrix}
\end{equation}
and $\mathbb{M}_{\chi \eta}$ is defined from Eq.(\ref{Eq:meta}). In order to build an invertible matrix $\mathbb{D}$ for the diagonalization of $\mathbb{M}_\xi$, we can fill it with the rotation matrices as presented in Eq.(\ref{Eq:Retachi}) and Eq.(\ref{Eq:RS}). Finally, from our previous discussion, by applying a rotation for the total mass matrix $\mathbb{M} = \mathbb{M}_{\xi_V} + \mathbb{M}_V$ through $\mathbb{D}$ we obtain a block-diagonal matrix, here represented by $4 \times 4$, that can be fully diagonalized after the insertion of an additional $\overline{\mathbb{D}}$.

\subsection{Self-Interactions of Gauge Bosons}\label{Sec:A8} 
The self-interactions of the gauge-bosons are mediated by
\begin{equation}\label{Eq:gb}
\mathcal{L}_{g.b.} = - \frac{1}{4} \mathbf{W}_{\mu \nu} \cdot \mathbf{W}^{\mu \nu} - \frac{1}{4} W^0_{\mu \nu} W_0^{\mu \nu}
\end{equation}
where $\mathbf{W}_{\mu \nu} = \left(W^1_{\mu \nu}, W^2_{\mu \nu}, \cdots, W^8_{\mu \nu}\right) $, and
\begin{equation}
W^a_{\mu \nu} = \partial_\mu W^a_{\nu} - \partial_\nu W^a_{\mu} + g f^{aki} W_\mu^k W_\nu^i
\end{equation}
or, in a simplified notation,
\begin{equation}
W^a_{\mu \nu} = \partial_\mu W^a_{\nu} - \partial_\nu W^a_{\mu} - g \text{Tr}\left[\mathbb{F}^a \cdot \mathbb{W}_{\mu \nu}\right] 
\end{equation}
with $(\mathbb{F}^a)_{ij} = f^{aij}$ the structure constants, and $(\mathbb{W}^{\mu \nu})_{ij} \equiv W^\mu_i W^\nu_j$.

Before perform the sum in Eq.(\ref{Eq:gb}), we first apply a set of transformations to the non-diagonal fields:
\begin{equation}
\begin{pmatrix}
W^4 \ (W^6) \\ W^5 \ (W^7)
\end{pmatrix} = \frac{1}{\sqrt{2}} 
\begin{pmatrix}
1 & 1 \\ -i & i
\end{pmatrix}
\begin{pmatrix}
V^{+q_V} \ (U^{+q_U}) \\ V^{-q_V} \ (U^{-q_U})
\end{pmatrix}
\end{equation}

\begin{equation}
\begin{pmatrix}
W^1 \\ W^2
\end{pmatrix} = \frac{1}{\sqrt{2}} 
\begin{pmatrix}
1 & 1 \\ i & -i
\end{pmatrix}
\begin{pmatrix}
W^+ \\ W^{-}
\end{pmatrix}
\end{equation}
and, finally,
\begin{equation}
\begin{pmatrix}
W^0 \\ W^3 \\ W^8
\end{pmatrix} = 
\begin{pmatrix}
- s_x & c_x c_w & -c_x s_w \\ 0 & s_w & c_w \\ c_x & s_x c_w & -s_x s_w
\end{pmatrix}
\begin{pmatrix}
Z' \\ A \\ Z
\end{pmatrix}
\end{equation}

One important feature of $\mathcal{L}_{g.b.}$ is the absence of $Z'$ interactions with standard vectors at tree- and loop-level. On the other hand, the charged bosons will couple with SM at loop-level and, for illustration, we present the vertices for $VW$ and the triple $VZA$:
\begin{equation}
V W : \ \mathcal{L}_{g.b.} \supset - \frac{g^2}{2} \left(V_\nu^- V^{+\nu}W_\mu^- W^{+\mu} + V_\mu^- W^{-\mu} V_\nu^{+} W^{+\nu} -2 V_\mu^+ W^{-\mu} V_\nu^{-} W^{+\nu} \right)
\end{equation}
\begin{eqnarray}
V Z A : \ \mathcal{L}_{g.b.} \  \supset& \ \frac{g^2}{4} \left(\sqrt{3} s_x (1 - 2s^2_w) + c_w s_w (1 - 3 s_x^2)\right) \times \nonumber \\
& \times \left(V_\nu^- A^\nu V_\mu^+ Z^\mu + V_\nu^+ A^\nu V_\mu^- Z^\mu - 
2 A^\nu Z^\nu V_\mu^+ V^{-\mu} \right)
\end{eqnarray}
Additional Feynman rules can be extracted from \cite{Cao:2016uur}.

\section{Fermions in the 3-3-1HL}\label{Sec:B}
We move to a brief description of the fermions in the 3-3-1HL. Among four possible versions, those models with $|\beta| = \frac{1}{\sqrt{3}}$ might require a special attention. This because they introduce heavy quarks and leptons with the same electric charges as the standard particles (see Eq.(\ref{Eq:qcharge}) - Eq.(\ref{Eq:lcharge})), corresponding a new pattern of mixing among different generations and then to a large set of original vertices at leading order \cite{Pleitez:1994pu}. We can summarize their fermion content like:
\begin{itemize}
	\item $\beta = - \frac{1}{\sqrt{3}}: \begin{cases} 
	\text{ - Three additional neutral heavy leptons (or right-handed neutrinos);}  \\
	\text{ - Two additional flavors for D quarks;} \\
	\text{ - One new flavor for U quarks.}
	\end{cases}$
	\item $\beta = \frac{1}{\sqrt{3}}: \begin{cases} 
	\text{ - Three additional heavy leptons with the electron charge;} \\
	\text{ - Two additional flavors for U quarks;} \\
	\text{ - One new flavor for D quarks.}
	\end{cases}$
\end{itemize}

\subsection{Gauge interactions of the Fermions}\label{Sec:B1}
We introduce the gauge interactions for fermions and, in a short notation, represent their left-handed triplets like
\begin{equation}\label{Eq:Fermdef}
\psi^L_\alpha = \begin{pmatrix}
\mathbf{L}^\alpha & E^\alpha
\end{pmatrix}_L; \qquad 
Q^L_i = \begin{pmatrix}
\mathbf{q}_i & J_i
\end{pmatrix}_L; \qquad 
Q^L_3 = \begin{pmatrix}
\mathbf{q}_3 & J_3
\end{pmatrix}_L
\end{equation}
where the index $\alpha = \left[e, \mu, \tau\right]$, $i = \left[1,2\right]$, and the boldface is just indicating the separation between SM doublets and exotic singlets. The right-handed fields follow the usual notation for the SM, i.e. $q^R_a, l^R_\alpha$ for quarks and leptons, $J^R_a, E^R_\alpha$ for the new particles. Thus, the total gauge-kinetic Lagrangian can be written as
\begin{eqnarray}\label{Eq:kin}
\mathcal{L}_{\text{kin}} &=& i \biggl\{ \sum_{\alpha = e, \mu, \tau} \overline{\psi}_\alpha^L \slashed D \psi_\alpha^L + \sum_{i = 1,2} \overline{Q}_i^L \slashed D Q_i^L + \overline{Q}_3^L \slashed D Q_3^L +  \nonumber \\ 
& & \sum_{a = 1,2,3} \overline{d}_a^R \slashed D d_a^R + \sum_{a = 1,2,3} \overline{u}_a^R \slashed D u_a^R + \sum_{i = 1,2} \overline{J}_i^R \slashed D J_i^R + \overline{J}_3^R \slashed D J_3^R\nonumber \\ 
& &  \sum_{\alpha = e, \mu, \tau} \overline{l}_\alpha^R \slashed D l_\alpha^R + \sum_{\alpha = e, \mu, \tau} \overline{E}_\alpha^R \slashed D E_\alpha^R \biggr\} 
\end{eqnarray} 
and each field associates its correspondent covariant derivative.

The following identity might be useful during the task of dividing the 3-3-1 interactions into SM and New Physics::
\begin{equation}\label{Eq:exp}
\begin{pmatrix}
\mathbf{a}^\dagger & b^*
\end{pmatrix}
\begin{pmatrix}
\mathbb{A} & \mathbf{x} \\
\mathbf{x}^\dagger & z
\end{pmatrix}
\begin{pmatrix}
\mathbf{a} \\
b
\end{pmatrix} = \mathbf{a}^\dagger \mathbb{A} \mathbf{a} + \mathbf{a}^\dagger \mathbf{x} b + b^* \mathbf{x}^\dagger \mathbf{a} + b^* z b
\end{equation}
such that the $2 \times 2$ matrix $\mathbb{A}$ for the covariant derivative will be given by
\begin{equation}
\mathbb{A} = D^{SM} + i \mathbf{g}^{Z'}_{2 \times 2}
\end{equation}
Apart from that, we define a vector of gauge bosons
$$\mathbf{x}_\mu \propto \frac{g}{\sqrt{2}} \begin{pmatrix}
V_\mu & U_\mu
\end{pmatrix}^\intercal$$
where the proportionality corresponds to the complex number $i$, for both $\mathbf{x}$ and $\mathbf{x}^\dagger$. The $b$ component represents the new heavy degrees of freedom, like in the field definition of Eq.(\ref{Eq:Fermdef}). By applying Eq.(\ref{Eq:exp}) in Eq.(\ref{Eq:kin}) we find
\begin{eqnarray}\label{Eq:kin2}
\mathcal{L}_{\text{kin}} = i \biggl\{ \overline{\mathbf{L}}_\alpha^L \slashed D \mathbf{L}_\alpha^L +  \overline{\mathbf{q}}_a^L \slashed D \mathbf{q}_a^L + \overline{l}_\alpha^R \slashed D l_\alpha^R  + \overline{d}_a^R \slashed D d_a^R  + \overline{u}_a^R \slashed D u_a^R \biggr\} + \nonumber \\ 
 + i \biggl\{ \overline{\mathbf{L}}_\alpha^L i \mathbf{g}_{Z'} \slashed{Z'} \mathbf{L}_\alpha^L +  \overline{\mathbf{L}}_\alpha^L \slashed{\mathbf{x}} E_\alpha^L + \overline{E}_\alpha^L \slashed{\mathbf{x}}^\dagger \mathbf{L}_\alpha^L  + \overline{E}_\alpha^L (\slashed{\partial} + i g_{Z'} \slashed{Z'} + i g_Z \slashed{Z} + i e Q \slashed{A}) E_\alpha^L \biggr\} + \nonumber \\ 
 + i \biggl\{ - \overline{\mathbf{q}}_i^L i \mathbf{g}^*_{Z'} \slashed{Z'} \mathbf{q}_i^L +  \overline{\mathbf{q}}_i^L \slashed{\mathbf{x}}^* J_i^L + \overline{J}_i^L \slashed{\mathbf{x}}^\intercal \mathbf{q}_i^L  + \overline{J}_i^L (\slashed{\partial} - i g^*_{Z'} \slashed{Z'} - i g^*_Z \slashed{Z} - i e Q \slashed{A}) J_i^L \biggr\} + \nonumber \\ 
+ i \biggl\{ \overline{\mathbf{q}}_3^L i \mathbf{g}_{Z'} \slashed{Z'} \mathbf{q}_3^L +  \overline{\mathbf{q}}_3^L \slashed{\mathbf{x}} J_3^L + \overline{J}_3^L \slashed{\mathbf{x}}^\dagger \mathbf{q}_3^L  + \overline{J}_3^L (\slashed{\partial} + i g_{Z'} \slashed{Z'} + i g_Z \slashed{Z} + i e Q \slashed{A}) J_3^L \biggr\} + \nonumber \\ 
 + i \biggl\{ \overline{E}_\alpha^R \slashed D E_\alpha^R +  \overline{J}_i^R \slashed D J_i^R + \overline{J}_3^R \slashed D J_3^R\biggr\} 
\end{eqnarray}
and note:
\begin{itemize}
	\item The first bracket corresponds to the SM gauge kinetic piece and we have just ignored the meaningless difference in this piece due to the conjugate representation. Thus, all the indices are running through the total three generations;
	\item The second contains the interactions for the heavy leptons, in addition to a tree-level term with $Z'$;
	\item The third contains new exotic quarks and the $Z'$ at tree-level. After the rotation of SM fields to their mass eigenstates, the matrix components of $\mathbf{g}_{Z'}$ give rise to flavor changing currents due to a distinct coupling with the third generation. The couplings must take into account the properties of the Gell-Mann matrices under conjugation and indices cover the first two generations;
	\item The fourth bracket, in the fundamental representation, is similar to the previous line;
	\item The last bracket contains the right-handed exotic fields. The leptonic indices cover the three generations while the following term contains the first two heavy quarks;
	\item The breaking of $\NM$ does not lead exactly to a theory invariant under $\SM$, unless the mass of the vectors  inside the doublet $\mathbf{x}$ are equal, valid only in the limit $u \gg v_\eta, v_\rho$.
\end{itemize}
Moreover, the next section discuss the presence of $\beta$-dependent Yukawas that mix standard and exotic fermions, thus expanding the first line into additional tree-level contributions. The Eq.(\ref{Eq:kin2}) can be symbolically written as
\begin{equation}
\mathcal{L}_{\text{kin}} = \mathcal{L}^{SM}_{\text{kin}} + \mathcal{L}^{E_L}_{\text{kin}} + \mathcal{L}^{J_{\alpha L}}_{\text{kin}} + \mathcal{L}^{J_{3 L}}_{\text{kin}} + 
\mathcal{L}^R_{\text{kin}}
\end{equation} 

We conclude this section with a brief digression on the described vertices. The present article composes the first part of a work in progress that intends to cover the complete integration of these new heavy fields. Thus, the method will result into an effective version of the Standard Model raised through the 3-3-1 gauge structure. Here we note, for example, that the only gauge-interactions generated at leading order are those with $Z'$. In the second part of this project we will compare the contributions of mixed terms involving exotic fields, loop-suppressed in the expansion of dimension-six operators. Thus, the complete set of terms involving $\mathbf{x}$ only contribute at this level, concealing the interactions with the non-diagonal gauge bosons. The one-loop sector of this Effective Theory will contain only the electromagnetic covariant derivative, i.e. the $U(1)$ invariant piece, apart from the interactions with $Z$. The terms with three exotic particles must be discarded. Finally, as we mentioned in the previous paragraph, for specific $\beta$-dependent interactions, the model suppression is reduced from the appearance of linear vertices on the heavy quarks.

\subsection{Yukawa Lagrangian}\label{Sec:B2}

Through this last section we discuss the division of Yukawa interactions into those pieces independent of the 3-3-1HL version and those only valid for a particular value of $\beta$. In the general case, the interactions are extracted from
\begin{eqnarray}\label{Eq:Yukagen}
\mathcal{L}_Y &=& \Bigl(\lambda^d_{i,a} \ \overline{Q}^i_L \eta^* d^a_R + \lambda^d_{3,a} \ \overline{Q}^3_L \rho d^a_R + 
\lambda^u_{i,a} \ \overline{Q}^i_L \rho^* u^a_R + \nonumber \\
& & + \lambda^u_{3,a} \ \overline{Q}^3_L \eta u^a_R + 
\lambda^j_{i,k} \ \overline{Q}^i_L \chi^* J^k_R + 
\lambda^j_{3,3} \ \overline{Q}^3_L \chi J^3_R + \nonumber \\
& & + \lambda^{d*}_{i,a} \ \overline{d}^a_R \eta^\intercal Q^i_L + 
\lambda^{d*}_{3,a} \ \overline{d}^a_R \rho^\dagger Q^3_L + 
\lambda^{u*}_{i,a} \ \overline{u}^a_R \rho^\intercal Q^i_L + \nonumber \\
& & + \lambda^{u*}_{3,a} \ \overline{u}^a_R \eta^\dagger Q^3_L + 
\lambda^{J*}_{i,k} \ \overline{J}^k_R \chi^\intercal Q^i_L + 
\lambda^{J*}_{3,3} \ \overline{J}^3_R \chi^\dagger Q^3_L \Bigr) + \nonumber \\
& & \Bigl(\lambda^l_{a,b} \ \overline{\psi}^a_L \rho l^b_R + \lambda^E_{a,b} \ \overline{\psi}^a_L \chi E^b_R + 
\lambda^{l*}_{a,b} \ \overline{l}^a_R \rho^\dagger \psi^a_L + \lambda^{E*}_{a,b} \ \overline{E}^b_R \chi^\dagger \psi^a_L \Bigr)
\end{eqnarray} 
where we have tried to unify the notation presented in \cite{Pisano:1991ee}, \cite{Buras:2012dp} and \cite{Fonseca:2016xsy}. The indices run as $a,b = 1,2,3$ and $i,k = 1,2$. On what follows we clarify the reason for changing the lepton notation.  

There are a few comments we can make from the above $\beta$-independent Yukawa. Note that both the heavy leptons and the new exotic quarks couples only with the first breaking triplet $\chi$. If the Scalar, Vector and Kinetic Lagrangian cannot connect the high sector with the SM at leading order, the forced absence of mixing between the $\chi$ components and standard scalars may compel the new physics to emerge always by pairs of exotic fields, leaving the heavy leptons stable.

One additional form to contour this feature is by considering the complete set of allowed Yukawa terms in the framework of specific versions. We start with leptons:

\begin{itemize}
	\item $\overline{\psi}_L\rho E_R$ and $\overline{\psi}_L\chi l_R$ : 
	
	The total hypercharge for these terms is $X = \frac{1}{2} - \beta \frac{\sqrt{3}}{2}$ and would be invariant for $\beta = \frac{1}{\sqrt{3}}$, i.e. in the version with a neutral $U$ gauge boson and where the heavy leptons have the same charge as the electron. The $\left(E e\right)$ mixing creates decay channels $E_\alpha \rightarrow SM$ via Eq.(\ref{Eq:kin2}).
	
	\item $\overline{\psi}_L\eta E_R$: 
	
	Since $X_\psi = X_\eta$, the total hypercharge is equal to the H.L. electric charge, or $X = q_E$, which is neutral for $\beta = - \frac{1}{\sqrt{3}}$. Apart from that, $V$ is the complex neutral gauge-boson and the SM portal comes from the mixing with neutrino, $\left(E \nu\right)$. 
\end{itemize}
Since both triplets are in the fundamental representation, terms with conjugated scalars are forbidden for the gauge symmetry. 

These additional lepton interactions appeared in the context of $|\beta| = \frac{1}{\sqrt{3}}$, where the new quarks have the same electric charges as the standard fermions. For $|\beta| = \sqrt{3}$, however, the quarks $J$ have exotic charges and cannot mix with the $U$ or $D$ type. Thus, whenever the $\beta$-dependent interactions are omitted from the potential, we may not expect new decay channels for the leptons in these type of 3-3-1 models. Since the Yukawa is the last component of our total Lagrangian, from Eq.(\ref{Eq:lepcharge}) these stable particles would be electrically charged with $-2$ or $+1$. 
	
Similarly, new vertices can also be extracted for the quarks and we finally classify the remaining $\mathcal{L}_Y$ components in terms of the $\beta$ sign:

\begin{itemize}
	\item $\beta = + \frac{1}{\sqrt{3}}$:
	\begin{eqnarray}
	\mathcal{L}_Y \ \supset& \ \lambda^{\chi u}_{i,a} \ \overline{Q}^i_L \chi^* u^a_R + 
	\lambda^{\chi d}_{3,a} \ \overline{Q}^3_L \chi d^a_R +  
	\lambda^{\rho J}_{i,k} \ \overline{Q}^i_L \rho^* J^k_R + \nonumber \\
	& \lambda^{\eta J}_{i,3} \ \overline{Q}^i_L \eta^* J^3_R + 
	\lambda^{\rho J}_{3,3} \ \overline{Q}^3_L \rho J^3_R + 
	\lambda^{\eta J}_{3,i} \ \overline{Q}^3_L \eta J^i_R + h.c.
	\end{eqnarray}
	
	\item $\beta = - \frac{1}{\sqrt{3}}$:
	\begin{eqnarray}\label{Eq.top}
	\mathcal{L}_Y \ \supset& \ \lambda^{\chi d}_{i,a} \ \overline{Q}^i_L \chi^* d^a_R + 
	\lambda^{\chi u}_{3,a} \ \overline{Q}^3_L \chi u^a_R +  
	\lambda^{\rho J}_{i,3} \ \overline{Q}^i_L \rho^* J^3_R + \nonumber \\
	&+ \lambda^{\eta J}_{i,k} \ \overline{Q}^i_L \eta^* J^k_R + 
	\lambda^{\rho J}_{3,i} \ \overline{Q}^3_L \rho J^i_R + 
	\lambda^{\eta J}_{3,3} \ \overline{Q}^3_L \eta J^3_R + h.c.
	\end{eqnarray}
	As before, the indices run as $a = 1,2,3$ and $i,k = 1,2$.
\end{itemize}
There is one important remark on the Yukawa components above - If we return to the high-energy scenario where the symmetry breaking is given exclusively by $\chi$, there will still be one generation of standard leptons and quarks that acquire mass from their mixing with the new heavy fields. In other words, the two first terms of Eq.(\ref{Eq.top}), for example, is breaking the $SU(3)_L \otimes U(1)_X$ directly into $U(1)_q$ leaving the bottom and the top-quark massive along with $J_i$ and $J_3$. On the basis $(u \ c \ t \ J_3)$ this feature can be illustrated as:
$$\begin{pmatrix}
& & & a_{14} \\  & \mathbb{O} & & a_{24} \\
& & & a_{34} \\ a_{14}^* & a_{24}^* &  a_{34}^* & a_{44}
\end{pmatrix} \rightarrow 
\begin{pmatrix}
0 & & & \\  & 0 & &  \\
& & m_t & \\  &  &  & m_J
\end{pmatrix}$$
Thus, the theory suggests that a mass hierarchy might be originated from the presence of at least two distinct breaking scales.

The splitting of $\mathcal{L}_Y$ into SM and new terms is almost trivial, in the sense that only the mass Lagrangian of U and D quarks will contribute to the standard part. Notwithstanding, the diagonalization matrices $V_L^U$ and $V_R^U$, in order to consent with our recent Higgs phenomenology, might strongly constrain the parameters of these new Higgs sector. In the universal 3-3-1HL, i.e. in the context of only $\beta$-independent processes, the exotic quarks $J_i$, for $i = [1,2]$, must mix through a similar pattern as the standard quarks.

\section{Conclusions}\label{Sec:C}
In our brief presentation of the 3-3-1HL components, we aimed to arrange a detailed separation between Standard terms and New Physics in order to prepare a model integration. In other words, we have focused on the task of select the totality of pieces that must compose a set of effective operators generated at tree- and loop-level. We stressed the importance of the variable $\beta$ to define a particle content and paid a special attention on an universal context defined by the $\beta$-independent vertices. Since, in principle, there is not a strong reason for a specific 3-3-1HL choice, we have claimed that the first step to test its original gauge-structure is by considering those processes present in all possible variants of the model. In this scenario the Yukawa sector, for example, is loop-supressed due to the exclusive presence of exotic mixed terms\footnote{Namely, by vertices in pairs of new fields.}.

The work intended to be complementary to the review section of \cite{Cao:2016uur} and we considered the most general scalar self-interactions by including new $\beta$-specific allowed terms, as present in \cite{Sanchez-Vega:2016dwe}. Finally, we concluded that the omission of these particular contributions is equivalent to assume a discrete symmetry for the variant where $|\beta|=\sqrt{3}$, leaving the theory in the presence of stable particles.

\ack

The author would like to thank E.R. Schmitz, B.L. Sánchez-Vega, Javier Fuentes-Martín and A. Pich for the important discussions and appreciates Vicente Pleitez for reviewing the manuscript. The work was financially supported by the National Counsel of Technological and Scientific Development (CNPq-Brazil) and by the Coordination for the Improvement of Higher Education Personnel (CAPES-Brazil). The author would like to express his gratitude for the hospitality of the Institute de Física Corpuscular (IFIC-UV) during the composition of this article.

\section*{References}
\bibliographystyle{unsrt}
\bibliography{thebiblio}

\end{document}